\documentclass[11pt]{article}
\usepackage{geometry}                
\usepackage{amsmath}
\usepackage{amscd}
\usepackage{times}
\usepackage{epsfig}
\usepackage{amsfonts,amsthm,eucal,amssymb,amscd}
\usepackage{color}
\usepackage{graphicx}
\usepackage{epstopdf}
\usepackage{authblk}
\title{\bf Resonances and 
antibound states of P\"oschl-Teller potential: Ladder operators and SUSY partners}
\author[a]{D. {\c C}evik}
\author[b]{M. Gadella}
\author[a]{{\c S}. Kuru}
\author[b]{J. Negro}
\affil[a]{\small Department of Physics, Faculty of Science, Ankara University, 06100 Ankara, Turkey}
\affil[b]{\small Departamento de F\'isica Te\'orica, At\'omica y \'Optica and IMUVA,
\\ Universidad de Valladolid, E-47011 Valladolid, Spain}

\begin{document}
\maketitle

\begin{abstract}
We analyze the one dimensional scattering produced by all variations of the P\"oschl-Teller potential, i.e., potential well, low and high barriers.  
We show that the P\"oschl-Teller well and low barrier potentials have no resonance poles, but an infinite number of simple poles along the imaginary axis corresponding to bound and antibound states. A quite different situation arises on the  P\"oschl-Teller  high barrier potential, which shows an infinite number of resonance poles and no other singularities.  We have obtained the explicit form of their associated Gamow states. We have also constructed ladder operators connecting wave functions for bound and antibound states as well as for resonance states. Finally, using wave functions of Gamow and antibound states in the factorization method, we construct some examples of supersymmetric partners of the P\"oschl-Teller Hamiltonian. 
\end{abstract}



%
%
%
%



\section{Introduction}
\label{}
One dimensional models in quantum mechanics are relevant as they may serve to test a wide range of quantum properties. They are also useful in the study of spherically symmetric three dimensional models. Although many studies of quantum one dimensional models have been addressed to the analysis of bound or scattering states, there are also a big number of works concerning unstable quantum states, which occur quite often in nature. Resonances can be identified with unstable quantum states, which are two equivalent manners for the description of the same reality \cite{BOHM,BB,BK,BEU}. In addition, there exists another type of not normalizable states with real negative energy called antibound states \cite{BOHM,NU,NAZAR,DRSL}. 


Resonances are defined as pairs of poles of the analytic continuation of the scattering  matrix ($S$ matrix). In the momentum representation, this analytic continuation is given by a meromorphic function $S(k)$ on the complex plane.  Then, resonance poles are symmetrically located on the lower half of the complex 
momentum plane  with respect to the imaginary axis. In the energy representation, the analytic continuation of the $S$ matrix is meromorphic on a two sheeted Riemann surface \cite{BOHM}. Now,  each pair of resonance poles are complex conjugated of each other with real part $E_R$ and imaginary part $\pm\Gamma/2$. These parameters $E_R$ and $\Gamma$ are the same that characterize a quantum unstable state: the resonant energy $E_R$ (which is the difference between the energy of the decaying state and the decay products) and the width $\Gamma$ (which is related with the inverse of the half life of the unstable state).   These states can also be represented by wave functions, which are eigenfunctions of the Hamiltonian with complex eigenvalue $E_R\pm i\,\Gamma/2$. Since the Hamiltonian is usually taken to be self adjoint on a Hilbert space, these wave functions called {\it Gamow states} (resonance states), can not be normalized \cite{BOHM,BG,CG}. 
There are some other definitions of resonances and quantum unstable states, not always equivalent, see references and  a brief review in \cite{AG}.  
In the present paper for practical reasons, we deal with resonances as pair of poles of the $S$ matrix in the momentum representation. For details concerning other formalisms we address to the literature in the subject  \cite{NU,FGR,KKH,RSIII,AJS,YAF}. 

In the study of the analytic properties of the $S$ matrix in the momentum representation  \cite{NU}, one sees the existence of three types of isolated singularities. One is the mentioned resonance poles, which may have multiplicity one or higher \cite{MON,MON1}. In addition, it may exist simple poles on the positive part of the imaginary axis, ($ik$, $k>0$) with energy $E=-{\hbar^2|k|^2}/{2m}$. Each one  determines the existence of one bound state $\psi$ and viceversa for each bound state there exists one of such poles. In this case $\psi$ is normalizable, i.~e., square integrable. 
Simple poles on the negative part of the imaginary axis
($ik$, $k<0$) correspond to other type of states called antibound or virtual states. 
Wave functions of antibound states are not square integrable; furthermore they blow up at the infinity. Their physical meaning is sometimes obscure (see \cite{BOHM,NAZAR} and references quoted therein). 
It is also possible the presence of a simple pole at the origin without physical meaning. 

These three types of states: resonance, bound and antibound, corresponding
to the singularities of the $S$ matrix, can also be obtained by imposing purely 
outgoing conditions to the solutions of the Schr\"odinger equation, as
we shall see in the next sections.

Coming back to resonance poles, we know that they come into pairs.  In the energy representation, these pairs are complex conjugate of each other, so that if $z_R=E_R-i\,\Gamma/2$ is one such pole, $z_R^*=E_R+i\,\Gamma/2$ is another one.   Then, for each resonance, there are two Gamow states, the so called {\it decaying Gamow state}, $\psi^D$, satisfying $H\psi^D=z_R\psi^D$ and the {\it growing Gamow state}, $\psi^G$ with $H\psi^G=z_R^*\psi^G$. The decaying Gamow state $\psi^D$ decays exponentially to the future,  i.e., $e^{-itH}\psi^D=e^{-iE_Rt}\,e^{-t\Gamma/2}\psi^D$, while the growing Gamow vector decays exponentially to the past (and grows exponentially to the future, hence its name), i.e., $e^{-itH}\psi^G=e^{-iE_Rt}\,e^{t\Gamma/2}\psi^G$. These formulas make sense in an appropriate rigged Hilbert space \cite{BG,CG}.
In the momentum representation,  poles on the forth quadrant correspond to decaying Gamow states and poles on the third quadrant correspond to growing Gamow states. 

In the present paper we shall deal with the hyperbolic 
P\"oschl-Teller  potential characterized by a parameter $\lambda$. 
Depending on the values of such parameter, this potential admits bound and antibound or even resonance states. Their corresponding poles of the $S$ matrix  will be determined analytically, contrarily to most of known resonance models where the poles have to be computed by numerical methods \cite{DRSL,AGML,VER1,VER2}. 

The factorization method has been used since the early times of quantum mechanics in order to obtain the spectrum of some Hamiltonians by algebraic means \cite{CHIS,BOG1}. 
For the hyperbolic P\"oschl-Teller potential, we may distinguish three different situations: potential well, low barrier and high barrier. In the first case, there exists bound and antibound states which are obtained from each other through ladder operators. In the second, there exist antibound states only, although the situation is similar to the former. Finally, the high barrier has an infinite number of resonances and the corresponding growing and decaying Gamow states are related by two different types of ladder operators. The ladder operators form an algebra supported by eigenfunctions of the Hamiltonian, which in general lie outside the Hilbert space. Up to now, this type of ladder operators were applied to bound states but they  have never been
applied to antibound states and resonances. This is a very
important result: 
resonance and antibound states share the same algebraic properties
as bound states.

Another application of the factorization method is to find Hamiltonian hierarchies starting from 
a given Hamiltonian, see \cite{CHIS} and references therein.
In order to obtain a Hamiltonian of a hierarchy, one uses in general an eigenfunction without zeros of the initial Hamiltonian, often corresponding to the ground state. However,  one rarely  uses an antibound state or a Gamow state to construct supersymmetric partners \cite{oscar1,oscar2,oscar3}. 
With these ideas in mind, we give some examples in which we build supersymmetric partners of  P\"oschl-Teller potentials using wave functions of antibound and  Gamow states. 

This paper is organized as follows: In the next section, we review some basic and important facts, concerning the hyperbolic P\"oschl-Teller potential and introduce the basic notation. In Section 3, we discuss some of its scattering properties. We show that all poles of the $S$ matrix corresponding to the purely outgoing boundary conditions, giving the energies and momenta of bound, antibound and resonance states, can be obtained analytically and exactly. This is a very exceptional outcome, since for the vast majority of worked potentials these poles can
only be obtained  by numerical methods. The construction of ladder operators relating eigenfunctions for bound and antibound states, growing and decaying Gamow states  is done in Section 4. In section 5, we get SUSY partners of the P\"oschl-Teller Hamiltonian, using antibound and Gamow states. We close our presentation with concluding remarks.

\section{The hyperbolic P\"oschl-Teller potential}

Let us consider the following one dimensional Hamiltonian:
\begin{equation}\label{1}
H=-\frac{\hbar^2}{2m}\,\frac{d^2}{dx^2}- \frac{\hbar^2}{2m}\, \frac{\alpha^2\,\lambda(\lambda-1)}{\cosh^2\alpha x}\,,
\end{equation}
where the second term in the right hand side of (\ref{1}) is known as the real hyperbolic P\"oschl-Teller potential. Here, $\alpha$ is a fixed constant while $\lambda$ is a parameter. 
Along the present paper, we shall consider three possibilities for $\lambda$ each one giving a different shape for the potential. They will be studied separately:
\begin{itemize}
 \item $\lambda>1$, potential well,
 \item $\frac{1}{2}\leq\lambda<1$, low barrier,
 \item $\lambda=\frac{1}{2}+i\ell; \,\,\ell>0$,  high barrier.
\end{itemize}
\begin{figure}
\centering
\includegraphics[width=0.4\textwidth]{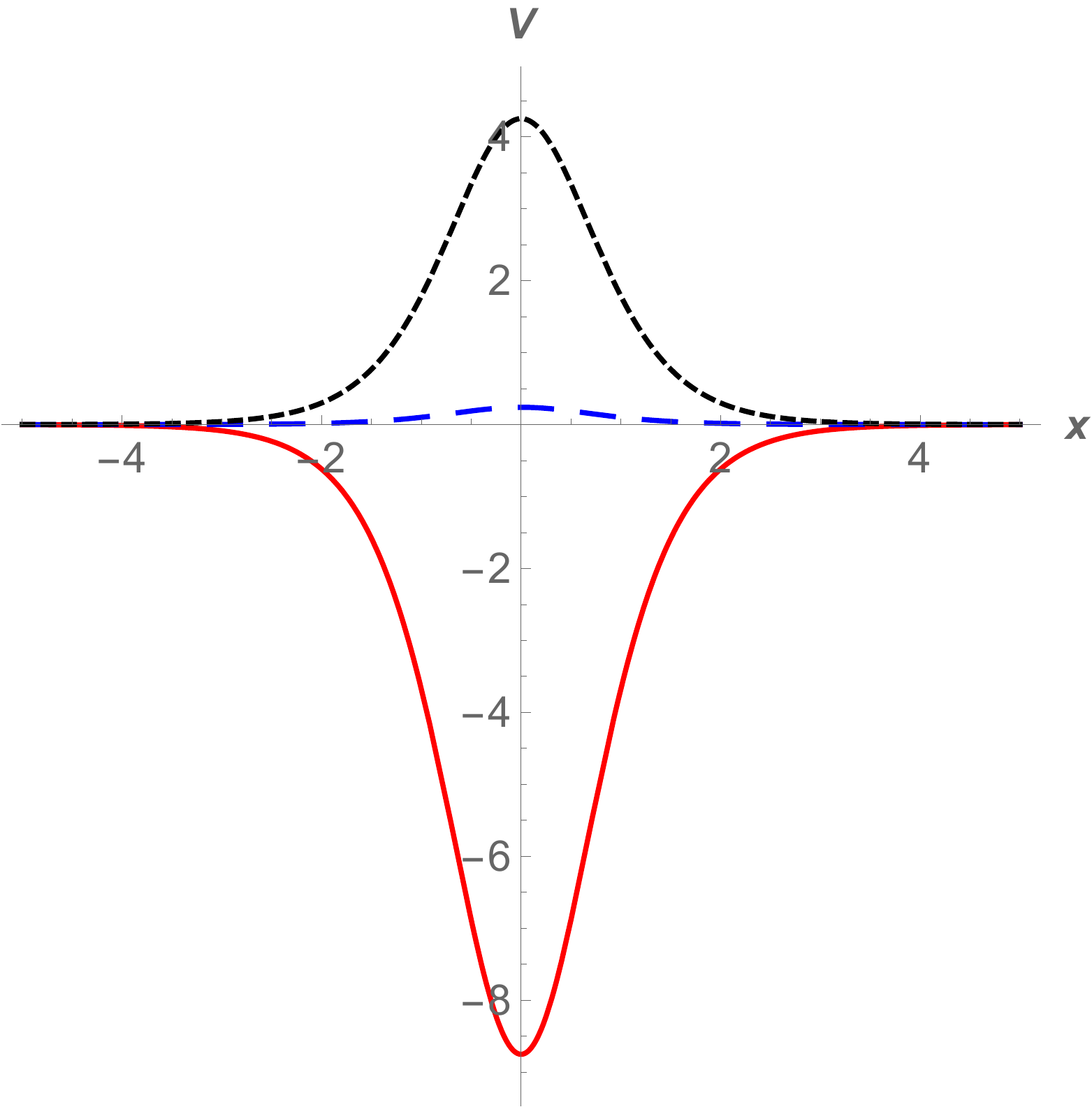}
\caption{\small Plot of the P\"oschl-Teller potential for different values of $\lambda$. The continuos line corresponds to  $\lambda=3.5$ (well), the dashing line to $\lambda=0.6$ (low barrier), the dotted line to $\lambda=1/2+i\,2$ (high barrier).
\label{potential}}
\end{figure}

The justification for the assigned names comes from their shapes shown in 
Fig.~\ref{potential}. Obviously, for the value $\lambda=1$ the potential vanishes. 
For integer values of $\lambda$, greater than one, it is well known that the resulting potential is reflectionless.

The time independent Schr\"odinger equation produced by the Hamiltonian (\ref{1}), has been widely studied \cite{Flugge,KN}. Nevertheless, the forthcoming presentation is quite relevant in order to follow our arguments. If we denote by $U(x)$ the wave function, the time independent Schr\"odinger equation obtained after the Hamiltonian (\ref{1}) is given by:

\begin{equation}\label{2}
U''(x)+ \left[k^2+\frac{\alpha^2\lambda(\lambda-1)}{\cosh^2\alpha x}\right]U(x)=0\,,
\end{equation}
where $k^2=\dfrac{2mE}{\hbar^2}$ and $U''(x)=\dfrac{d^2U}{dx^2}$. Then, let us introduce the following new variable
\begin{equation}\label{3}
y(x):=\tanh\alpha x\,
\end{equation}
and the new function $\nu(y)$, 
\begin{equation}\label{5}
U(y)=(1+y)^r(1-y)^s\nu (y)\,,
\end{equation}
where $r$ and $s$ are 
\begin{equation}\label{7}
r=\frac{ik}{2\alpha}\,, \qquad s =-\frac{ik}{2\alpha}\,.
\end{equation}
With these choices, equation (\ref{2}) becomes the Jacobi equation,
\begin{equation}\label{8}
(1-y^2)\,\nu''(y)+\left[\frac{2ik}{\alpha}-2y\right]\nu'(y)+\lambda(\lambda-1)\, \nu(y)=0\,.
\end{equation}
In order to write (\ref{8}) in the standard form of the hypergeometric equation, we need to use  the following change of variable
\begin{equation}\label{9}
z:=\frac{y+1}{2}\,,
\end{equation}
so that (\ref{8}) takes the form:
\begin{equation}\label{10}
z(1-z)\,\nu''(z)+\left[\frac{ik}{\alpha}-2z+1\right]\nu'(z)+\left[\lambda(\lambda-1)\right]\nu(z)=0\,.
\end{equation}
Note that $\nu'(z)$ denotes $d\nu/dz$, etc.  We have finally reached the hypergeometric equation $z(1-z)\,\nu''(z)+[c-(a+b+1)z]\nu'(z)-ab\nu(z)=0$, with $a=\lambda$, $b=1-\lambda$ and $c=ik/\alpha+1$. Two independent solutions are given in terms of  hypergeometric functions: $_2F_1(a,b;c;z)$ and $z^{1-c}\,_2F_1(a-c+1,b-c+1;2-c;z)$ (provided that $c$ is not an integer \cite{AS}.) 
Therefore, the general solution of equation (\ref{2}) can be reached after some evident manipulations and is given by:
\begin{eqnarray}\label{11}
&&U(x)= A\,(1+\tanh\alpha x)^{ik/2\alpha}(1-\tanh\alpha x)^{-ik/2\alpha}{}_{2}F_{1}\left(\lambda,1-\lambda;\frac{ik}{\alpha}+1;\frac{1+\tanh\alpha x}{2}\right) 
\nonumber\\[2ex] 
&&\quad \quad+  B\,2^{ik/\alpha}(1+\tanh\alpha x)^
{-ik/2\alpha}(1-\tanh\alpha x)^{-ik/2\alpha} 
\nonumber\\[2ex] 
&&\quad \quad\times{}_{2}F_{1}\left(\lambda-\frac{ik}{\alpha},1-\lambda-\frac{ik}{\alpha};1-\frac{ik}{\alpha};\frac{1+\tanh\alpha x}{2}\right)\,,
\end{eqnarray}
where $A$ and $B$ are arbitrary constants.

The $S$ matrix connects the asymptotic forms of the 
incoming wave function with outgoing wave function. 
Fortunately the asymptotic behavior of the hypergeometric functions is well known \cite{AS} and the asymptotic form of (\ref{11}) is:
\begin{itemize}
\item
For $x\longmapsto+\infty$
\begin{equation}\label{uplus}
\begin{array}{rl}
\!\!\!\!\!\!\!\!\!\!\!\!\!\!\!\!\!\!\!U^+(x)&= \displaystyle
\left[ A\, \frac{\Gamma\left(\frac{ik}{\alpha}+1\right)\,\Gamma\left(\frac{ik}{\alpha}\right)}{\Gamma\left(\frac{ik}{\alpha}+1-\lambda\right)\, \Gamma\left(\frac{ik}{\alpha}+\lambda\right)} +B\, \frac{\Gamma\left(1-\frac{ik}{\alpha}\right)\, \Gamma\left(\frac{ik}{\alpha}\right)}{\Gamma\left(1-\lambda\right)\Gamma\left(\lambda\right)}  \right]\,e^{ikx} 
\\[2.75ex]
&+ \displaystyle\left[ A\,  \frac{\Gamma\left(\frac{ik}{\alpha}+1\right)\,\Gamma\left(-\frac{ik}{\alpha}\right)}{\Gamma\left(1-\lambda\right)\Gamma\left(\lambda\right)}  +B\, \frac{\Gamma\left(1-\frac{ik}{\alpha}\right)\, \Gamma\left(-\frac{ik}{\alpha}\right)}{\Gamma\left(\lambda-\frac{ik}{\alpha}\right)\Gamma\left(1-\lambda-\frac{ik}{\alpha}\right)}   \right]\,e^{-ikx}\,
\\[3ex]
&=\displaystyle A'\,e^{ikx}+B'\,e^{-ikx}\,
\end{array}
\end{equation}
\item
For $x\longmapsto-\infty$
\begin{equation}\label{13}
\!\!\!\!\!\!\!\!\!\!\!\!\!\!\!\!\!\!\!U^-(x)= A\,e^{ikx}+B\,e^{-ikx}\, .
\end{equation}
\end{itemize}
 
We recall that $\alpha$ is a given positive constant. In the sequel, we shall fix $\alpha=1$ for simplicity. Then, we can define the $S$ matrix that relates the asymptotically  incoming wave function with the asymptotically outgoing wave function \cite{B}:
\begin{equation}\label{14}
\left( \begin{array}{c} B \\[2ex] A'        \end{array}   \right) = \left( \begin{array}{cc}     S_{11}  &  S_{12}  \\[2ex]  S_{21}  &  S_{22}    \end{array}   \right) \left( \begin{array}{c} A \\[2ex] B'        \end{array}   \right)\,.
\end{equation}
The matrix elements $S_{ij}$ of the $S$ matrix are usually written in terms of the elements $T_{ij}$ of the transfer matrix $T$ which relates the asymptotic wave functions in the negative infinity and in the positive infinity is defined as
\begin{equation}\label{15}
\left(\begin{array}{c}
A' \\[2ex] 
B' \end{array}\right) =  
\left(\begin{array}{cc}
T_{11} & T_{12} \\[2ex]
T_{21} & T_{22}
\end{array}
\right) \left(\begin{array}{c}
A \\[2ex]
B\end{array}\right)\,,
\end{equation}
in the following form:

\begin{equation}\label{16}
S= \frac{1}{T_{22}}\left(\begin{array}{cc}
-T_{21} & 1 \\[2.ex]
T_{11} T_{22}-T_{21}T_{12} & T_{12}
\end{array}
\right)\,.
\end{equation}

The explicit form of the transfer matrix $T$ \cite{Alhassid,Guerrero}, obtained 
from (\ref{uplus}), (\ref{13}) and (\ref{15}),
is the following:
\begin{equation}\label{17}
T=\left(\begin{array}{cc}
\dfrac{\Gamma\left(ik+1\right)\Gamma\left(ik\right)}{\Gamma\left(ik+1-\lambda\right)\Gamma\left(ik+\lambda\right)} & \dfrac{\Gamma\left(1-ik\right)\Gamma\left(ik\right)}{\Gamma\left(1-\lambda\right)\Gamma\left(\lambda\right)} 
\\[2.5ex]
\dfrac{\Gamma\left(ik+1\right)\Gamma\left(-ik\right)}{\Gamma\left(\lambda\right)\Gamma\left(1-\lambda\right)} & \dfrac{\Gamma\left(1-ik\right)\Gamma\left(-ik\right)}{\Gamma\left(\lambda-ik\right)\Gamma\left(1-\lambda-ik\right)} 
\end{array}
\right)\,.
\end{equation}
It is easy to check that in this case $\det T=1$ and $S\,S^\dagger=S^\dagger\,S=1$. 
Thus, we have obtained the explicit form of the $S$ matrix in the momentum representation, which will be henceforth denoted by $S(k)$. 

Now, we define the purely outgoing states of the Schr\"o-
dinger equation in this case,
as the solutions characterized by (\ref{uplus}) and (\ref{13}) such that:
$A=B'=0$. In other words, the asymptotic behavior consist in outgoing
waves to the right and to the left of the potential range. From 
the $T$ matrix equation (\ref{15}), $B'=T_{21}A+T_{22}B$.
Therefore, the values of $k$ satisfying the purely outgoing boundary conditions reduce to  the 
solutions of $T_{22}(k)=0$.
As it is seen from (\ref{16}), this equation characterizes the poles of $S(k)$ which
are related with purely outgoing states. 
For such values of $k$, according to (\ref{11}), the wave functions corresponding to outgoing states are given (up to a constant factor) by 
\begin{eqnarray}
&&U(x)=   
2^{ik/\alpha}(1+\tanh\alpha x)^
{-ik/2\alpha}(1-\tanh\alpha x)^{-ik/2\alpha} 
\nonumber\\[2ex] 
&&\quad \quad  \times\,
{}_{2}F_{1}\left(\lambda-\frac{ik}{\alpha},1-\lambda-\frac{ik}{\alpha};1-\frac{ik}{\alpha};\frac{1+\tanh\alpha x}{2}\right)\, .
 \label{outgoing}
\end{eqnarray}

%
\section{Three types of P\"oschl-Teller potentials}
Along this present section, we intend to analyze all kind of features that  emerge from a scattering analysis of the three types of hyperbolic P\"oschl-Teller potentials under our study. This includes scattering states, resonances, bound and antibound states. We shall follow the order beginning with the potential well, then low barrier to conclude with the high barrier. 
\subsection{Potential well ($\lambda>1$)}
\begin{figure}
\centering
\includegraphics[width=0.40\textwidth]{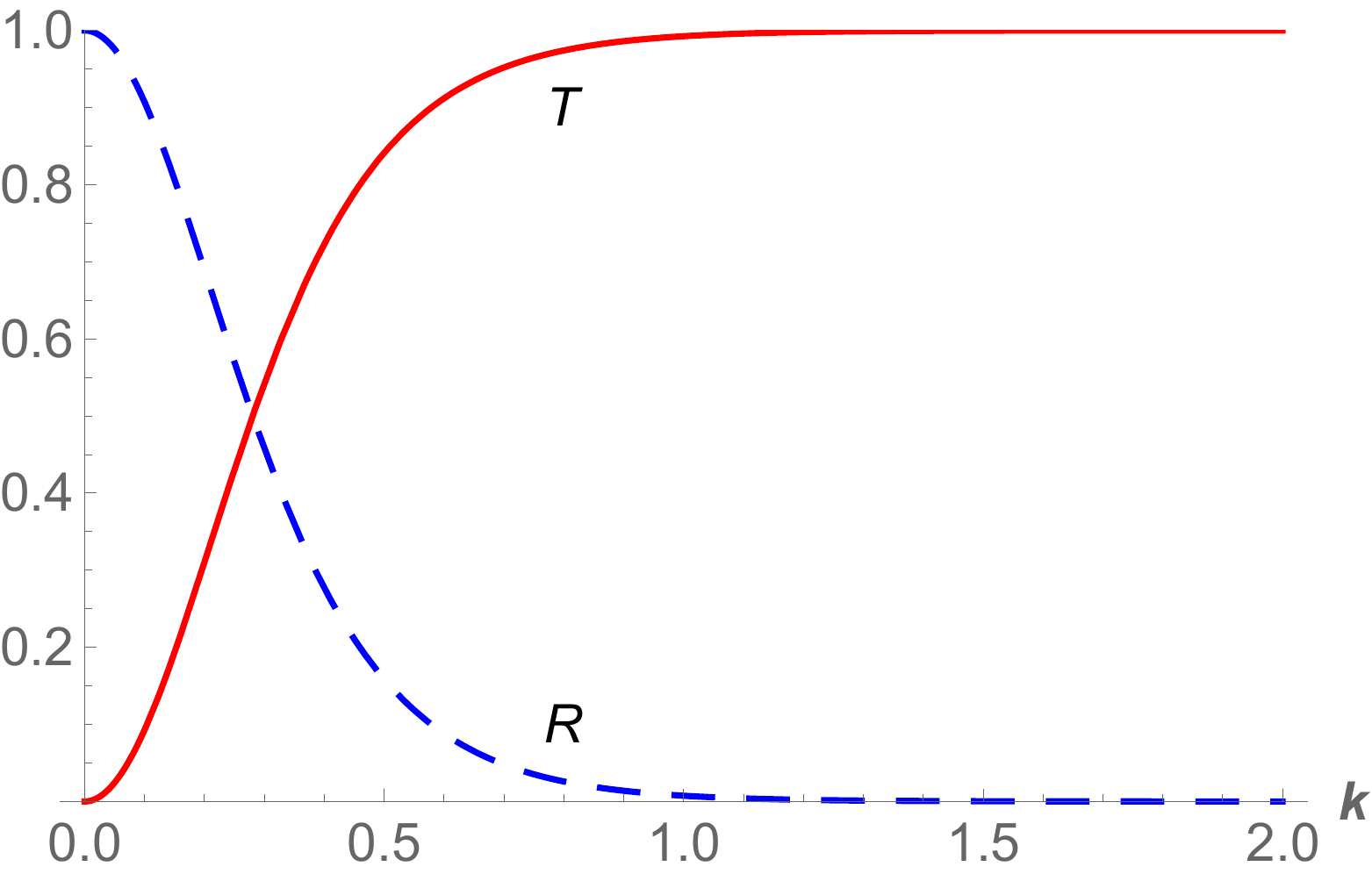}
\caption{\small Well potential: Plot of $T(k)$ and $R(k)$ for $\lambda=3.5$.}
\label{fig2}
\end{figure}

One of the most interesting objects in the study of scattering is the explicit forms of the reflection and transmission coefficients. The point of departure is now an asymptotic 
incoming plane wave from the left that after interaction with the potential
comes into a reflected and a transmitted plane waves characterized by $k \in \mathbb R$ .  This means that in  (\ref{uplus}) and (\ref{13}), we take $A=1$ and $B'=0$. Then, we obtain the following reflection $r$ and transmission $t$ amplitudes:
\begin{equation}
\begin{array}{l}
r= B= S_{11} = -\dfrac{T_{21}}{T_{22}} = 
\dfrac{\Gamma\left( {ik} \right)\Gamma\left(\lambda- {ik} \right)\Gamma\left(1-\lambda- {ik} \right)}{\Gamma\left(- {ik} \right)\Gamma\left(1-\lambda\right)\Gamma\left(\lambda\right)}\,,
\\[2.75ex]
t= A' = S_{21} = \dfrac1{T_{22}} = \dfrac{\Gamma\left(\lambda-{ik}\right)\Gamma\left(1-\lambda-{ik}\right)}{\Gamma\left(1-{ik}\right)\Gamma\left(-{ik}\right)}\,.
\end{array}
\end{equation}
Then, the reflection and the transmission coefficients are given by
\begin{equation}\label{18}
\begin{array}{l}
R=|r|^2=\left| \dfrac{\Gamma\left( {ik} \right)\Gamma\left(\lambda- {ik} \right)\Gamma\left(1-\lambda- {ik} \right)}{\Gamma\left(- {ik} \right)\Gamma\left(1-\lambda\right)\Gamma\left(\lambda\right)}\right|^2 \,,
\\[2.75ex]
T=|t|^2= \left|  \dfrac{\Gamma\left(\lambda-{ik}\right)\Gamma\left(1-\lambda-{ik}\right)}{\Gamma\left(1-{ik}\right)\Gamma\left(-{ik}\right)} \right|^2\,.
\end{array}
\end{equation}
We can check that $T+R=1$ for $k \in \mathbb R$.  In Fig.~\ref{fig2}, we plot $T(k)$ versus $R(k)$ for  
$\lambda=3.5$. We recall that when $\lambda$ is an integer, then the transmission coefficient is equal to one: $T=1$. Consequently, $R=0$ and we have a reflectionless potential. 

Now, consider that $k\in \mathbb C$. As was settled earlier, singularities of $S(k)$ corresponding to outgoing states, are determined via the equation $T_{22}(k)=0$, which in our case takes the form:
\begin{equation}\label{19}
\frac1{t(k)}=\frac{\Gamma(1-ik)\,\Gamma(-ik)}{\Gamma(\lambda-ik)\,\Gamma(1-\lambda-ik)}=0\,.
\end{equation}
\begin{figure}
\centering
\includegraphics[width=0.4\textwidth]{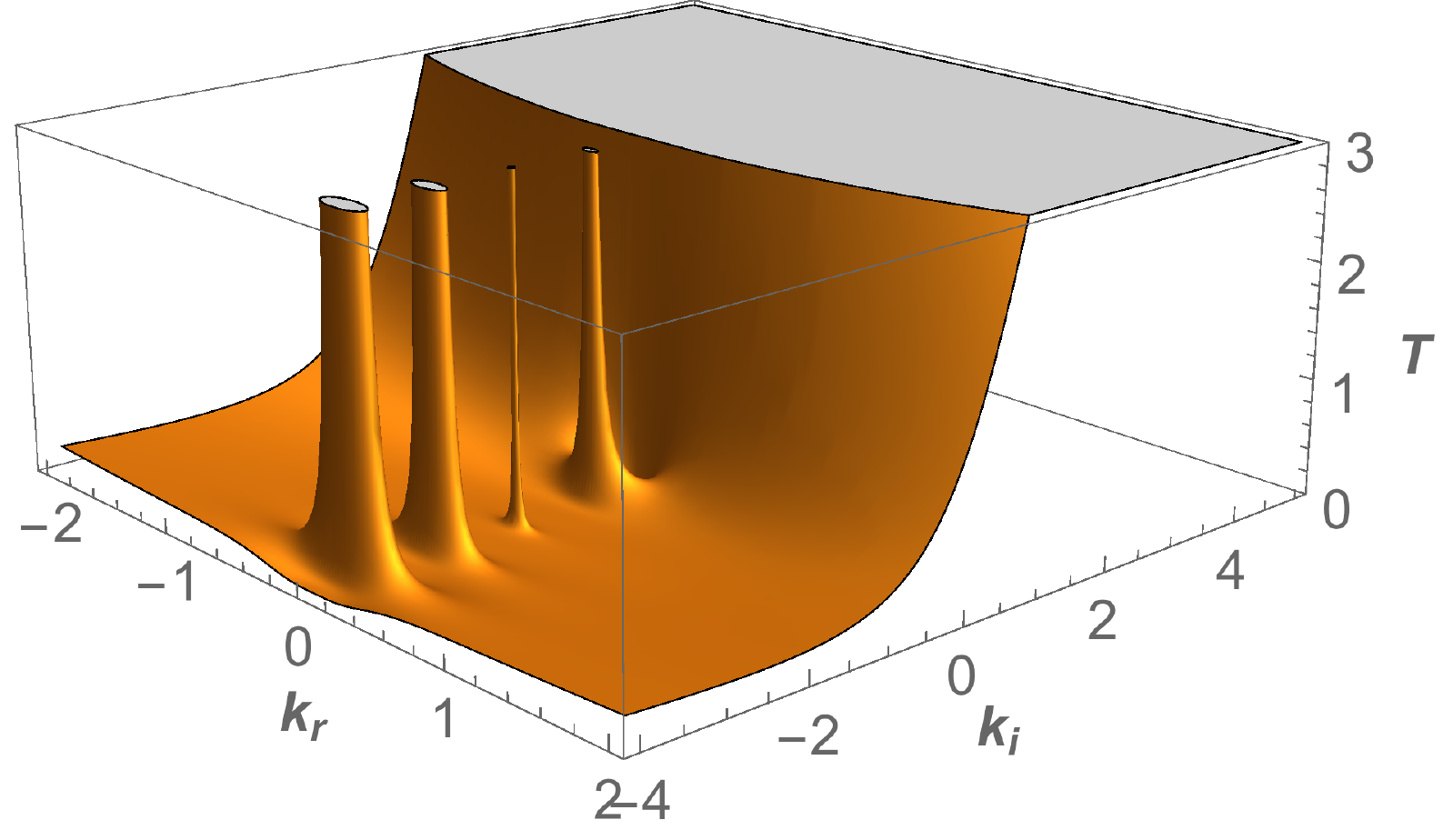}
\quad
\includegraphics[width=0.4\textwidth]{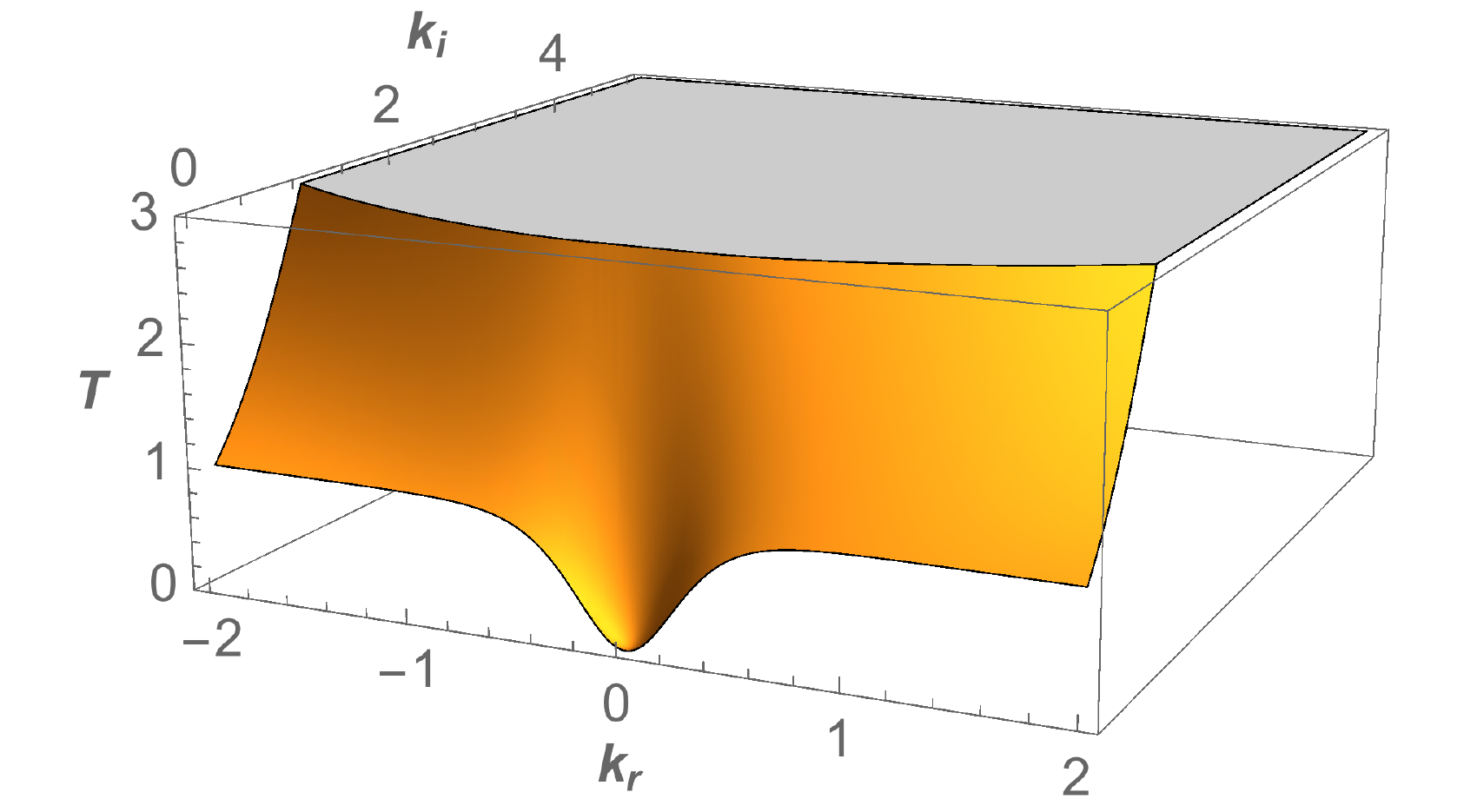}
\caption{\small Well potential: Plot of $T(k)$ for  
$\lambda=3.5$, and complex values $k= k_r+ i k_i$. The singularities are shown at $k_2(n): i\,2.5, i\, 1.5, i\, 0.5,
-i\, 0.5,-i\, 1.5,-i\, 2.5,-i\, 3.5$ (left). At the right, it is shown the profile of
$T(k)$ when $k_i=0$. This coincides with the transmission coefficient of Fig.~\ref{fig2}
(extended to $-\infty<k<+\infty$).
\label{figpoloswell}}
\end{figure}
\begin{figure}
\centering
\includegraphics[width=0.40\textwidth]{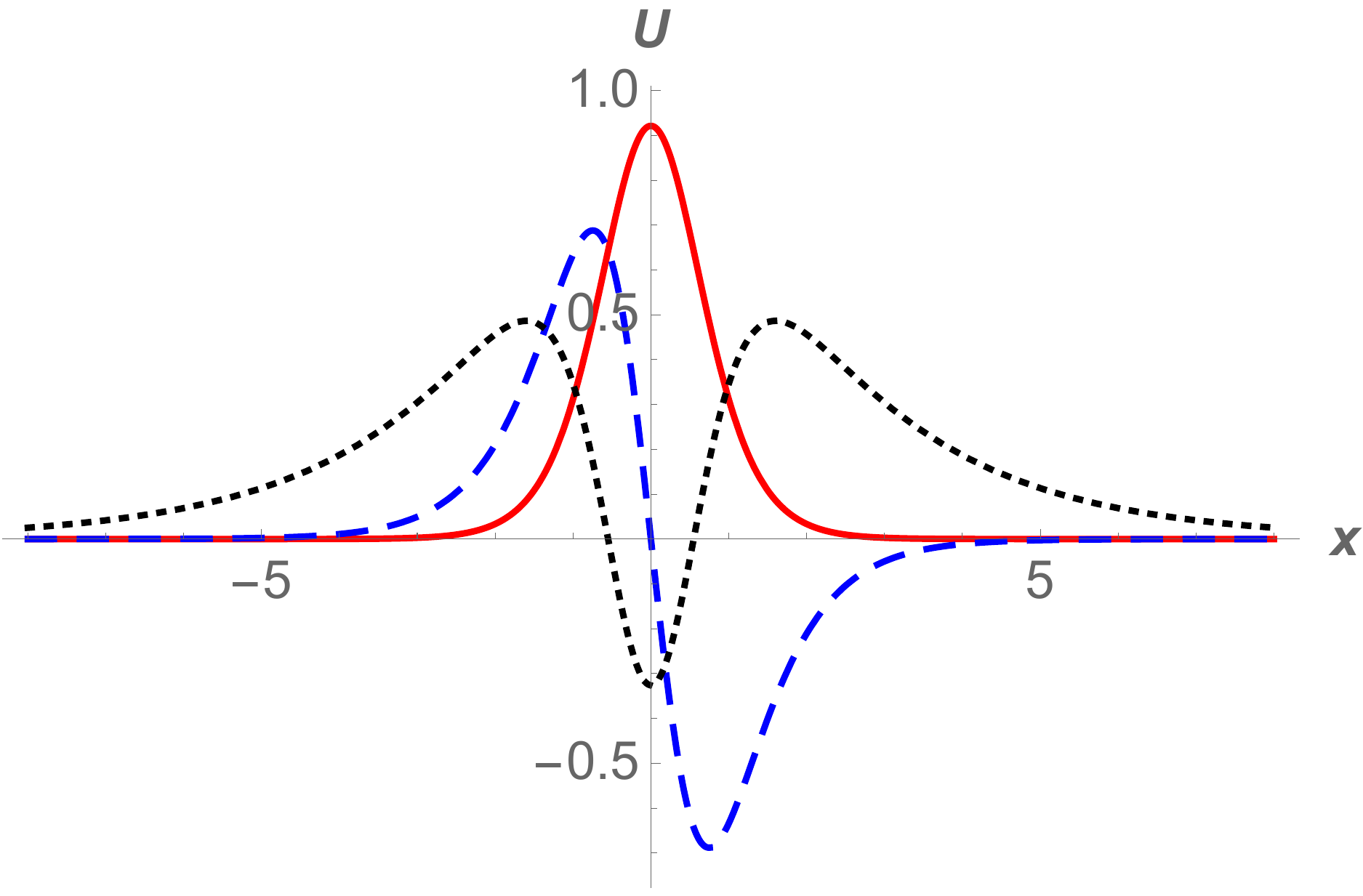}
\caption{\small Well potential: The plot of bound state wave functions with $\lambda=3.5$ and $n=0$ (continuous), $n=1$ (dotted) and $n=2$ (dashed). Its corresponding values of $k_2(n)$ are:  
$i 2.5$, $i 1.5$ and $i 0.5$. 
}\label{FIG_3}
\end{figure}
Therefore, such singularities of $S(k)$ coincide with the singularities of
$t(k)$ or $T(k)$.
Since the Gamma function has no zeros, the solutions of (\ref{19}) are restricted to the poles of the two Gamma functions in the denominator. 
Therefore, solutions of (\ref{19}) satisfy either $\lambda-ik=-n$ or $1-\lambda-ik=-n$, with $n=0,1,2,\dots$. If we call $k_1(n)$ and $k_2(n)$ to the solutions of the first and second type, respectively, we have that
\begin{equation}\label{20}
k_1(n)=-i(n+\lambda)\,,\qquad k_2(n)=-i(n-\lambda+1)\,.
\end{equation}

When $\lambda>1$, where $\lambda$ is not an integer, solutions $k_1(n)$ are all located in the negative part of the imaginary axis. Poles on the negative imaginary axis are called antibound poles. Their corresponding real energies are eigenvalues of the Hamiltonian and their respective eigenstates are called  antibound states. Wave functions for antibound states are not square integrable and diverge at the infinity. 
All this means that our potential shows an infinite number of equally spaced antibound poles.  Now, let us focus our attention in the second identity in (\ref{20}), $k_2(n)$.  The inequality $n-\lambda+1<0$  has at least one solution and the number of its solutions is always finite. Consequently, the solutions $k_2(n)$ give a finite number of poles in the positive imaginary semiaxis, which define bound states \cite{NU} and an infinite number of antibound poles. 

\begin{figure}
\centering
\includegraphics[width=0.4\textwidth]{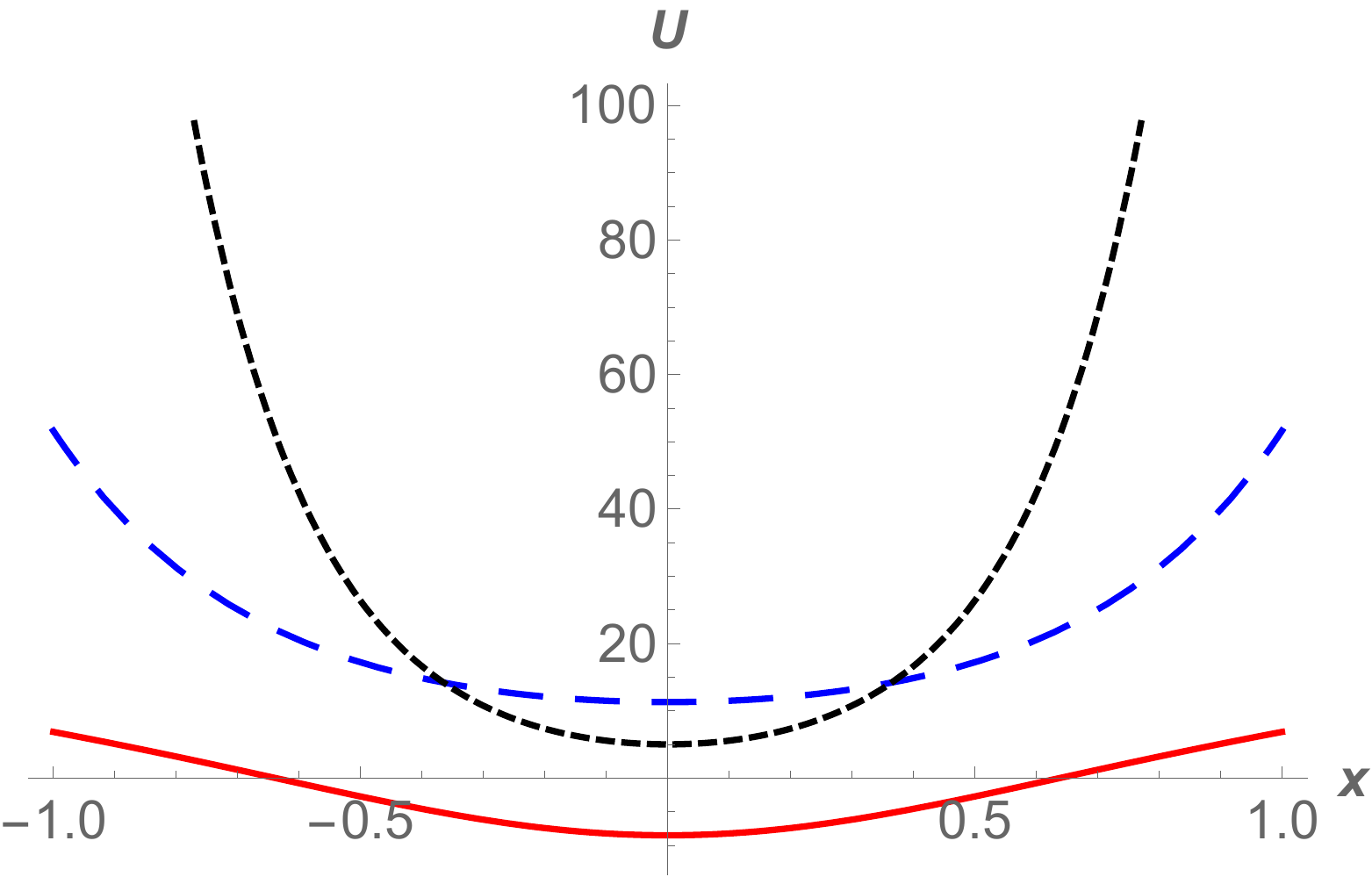}\qquad
\includegraphics[width=0.4\textwidth]{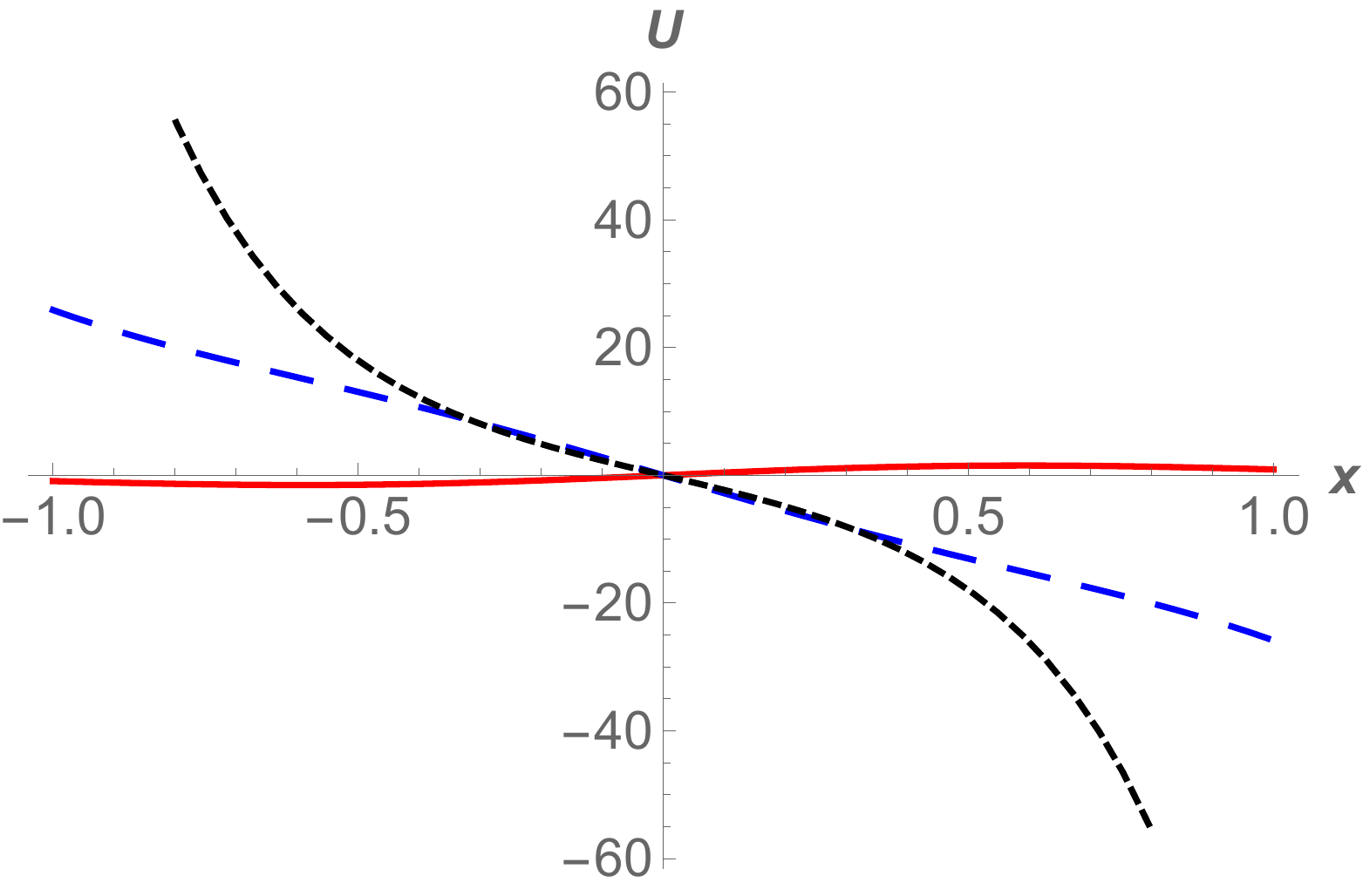}
\caption{\small Well potential: Plot of the antibound wave functions for the first three even values of 
$n$, 
$n=4$ (continuous line), $n=6$ (dashed line) and $n=8$ (dotted line) with $\lambda=3.5$ 
 (left). At the right, it is displayed the wave functions for the three first odd values of $n$, $n=3$ (continuous line), $n=5$ (dashed line) and $n=7$ (dotted line).
}\label{FIG_412}
\end{figure}

No resonances appear for these specific values of $\lambda$ ($\lambda>1$).
In Fig.~\ref{figpoloswell}, bound and antibound poles are shown in the plot of $T(k)$, $k\in \mathbb C$ (as mentioned above, such singularities
coincide with the poles of $T$). 
In Fig.~\ref{FIG_3}, we plot the first three bound state wave functions  and in Fig.~\ref{FIG_412} first six antibound state wave functions for the value  $\lambda=3.5$.
\subsection{Low barrier ($\frac12\le \lambda<1$)}

The transmission and the reflection coefficients are respectively given by:
\begin{equation}
\begin{array}{l}\label{21}
\displaystyle  T=\frac{\sinh^2(\pi k)}{\sin^2(\pi\lambda)+\sinh^2(\pi k)}\,,
\qquad
R=\frac{\sin^2(\pi\lambda)}{\sin^2(\pi\lambda)+\sinh^2(\pi k)}\,.
\end{array}
\end{equation}
Obviously, $T+R=1$ for $k \in \mathbb R$. 

\begin{figure}
\centering
\includegraphics[width=0.4\textwidth]{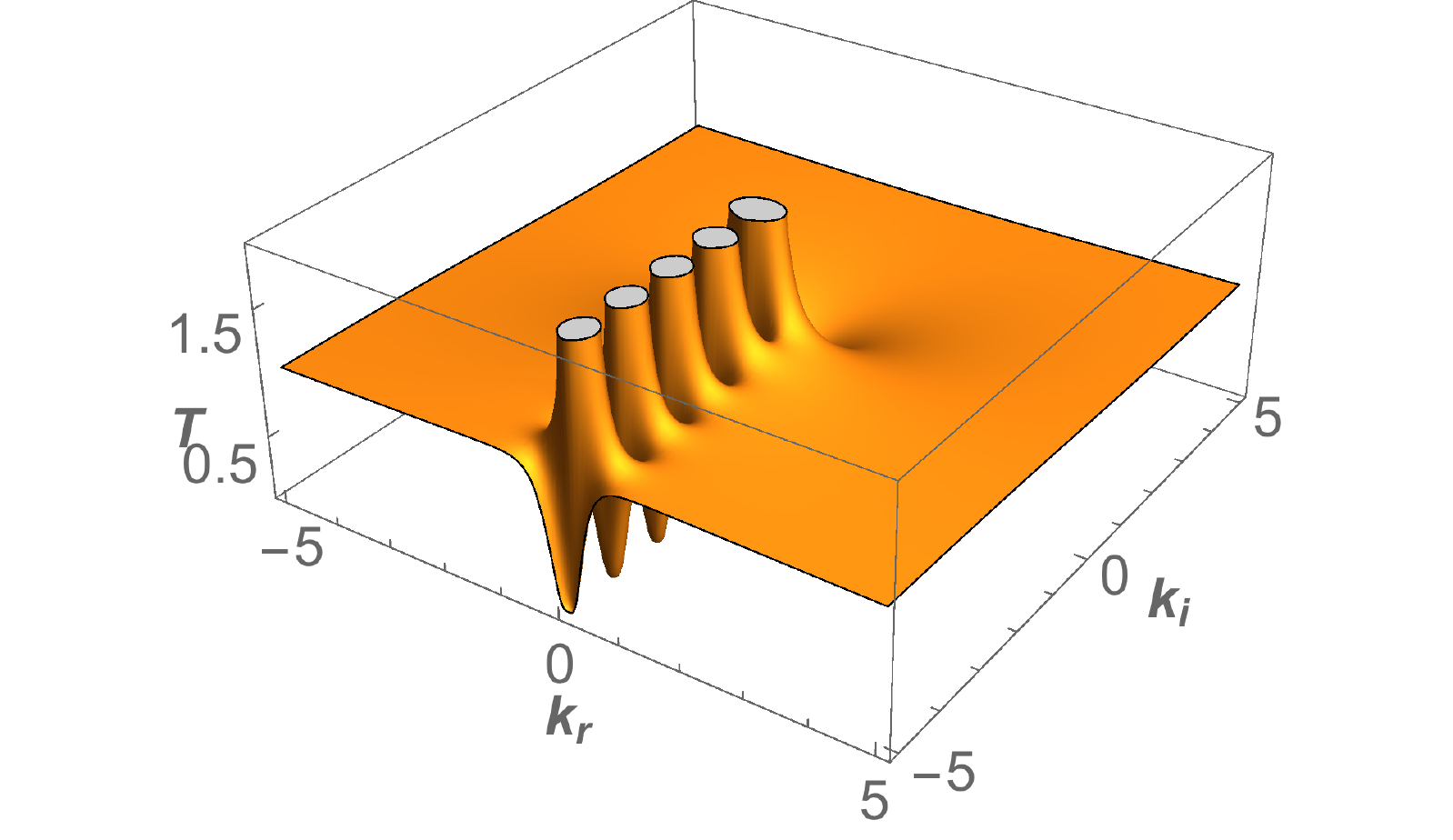}
\qquad
\includegraphics[width=0.4\textwidth]{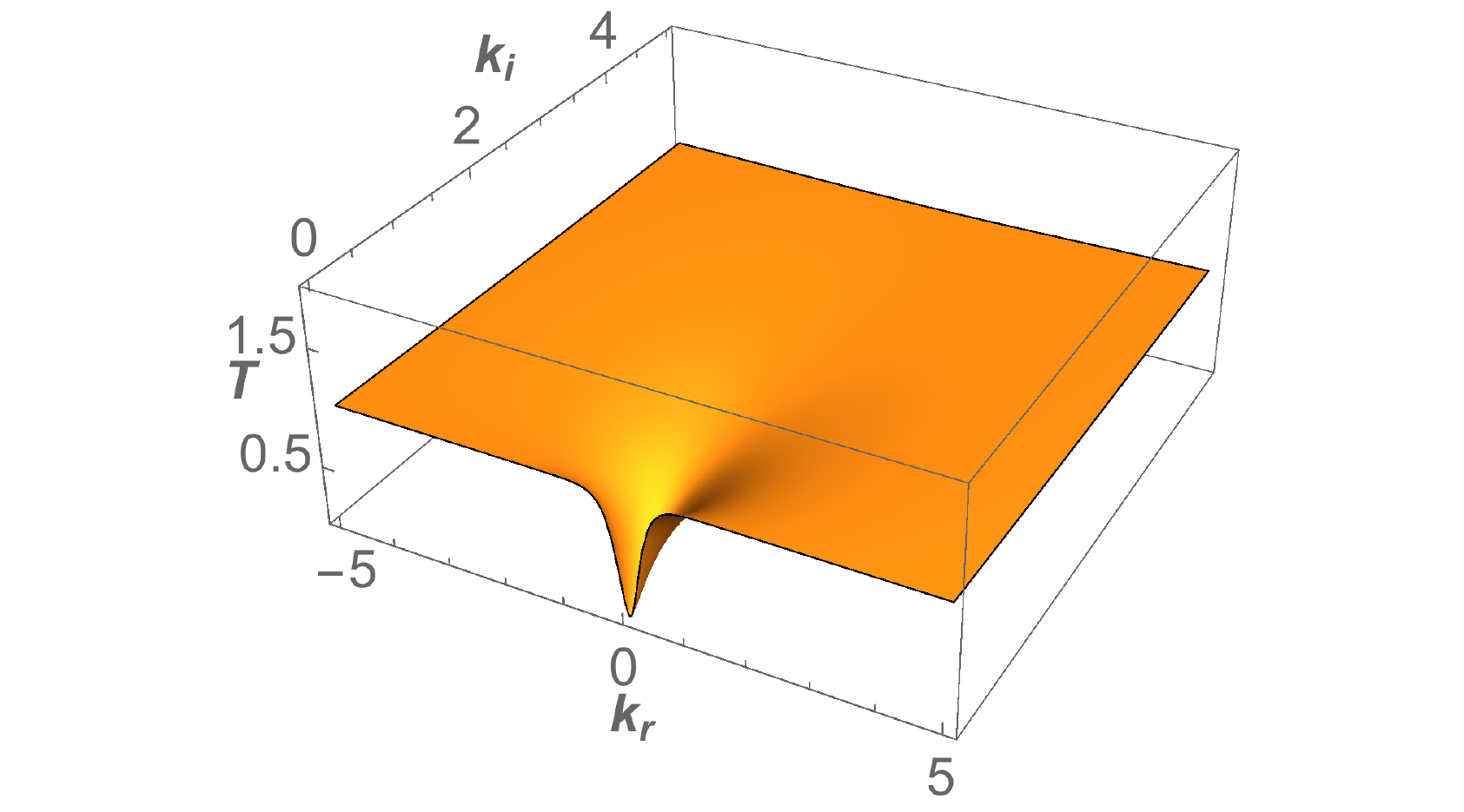}
\caption{\small Low barrier potential: Plot of $T(k)$ for  
$\lambda=0.75$ and complex values $k= k_r+ i k_i$. The singularities are shown at $k_2(n):  -i\, 0.25, -i\, 1.25, -i\, 2.25, -i\, 3.25, -i\, 4.25$ (left).
At the right, it is shown the profile of
$T(k)$ when $k_i=0$. This coincides with the shape of the transmission coefficient.
\label{polos_low}}
\end{figure}

In this case, the singularities of $S(k)$, $k\in \mathbb C$, are also given by equations (\ref{20}). However,  since $\frac12\le \lambda<1$, we always have that $n+\lambda>0$ and $n-\lambda+1>0$, so that no bound states exist here. Instead, we have two different series of antibound states,  where the antibound poles are given by $k_1(n)$ and $k_2(n)$ as in (\ref{20}). This is illustrated in Fig.~\ref{polos_low},  where $T(k)$, for $\lambda = 0.75$, is represented. 
The plot of $T(k)$, $R(k)$ and the shape of wave functions 
for antibound states are quite similar to the previous case.
\subsection{High barrier ($\lambda=\frac12+i\ell$)}

To start with, let us give the expressions for the transmission and reflection coefficients:
\begin{equation}
\begin{array}{l}\label{22}
\displaystyle 
T=\frac{\sinh^2(\pi k)}{\cosh^2(\pi k)+\sinh^2(\pi \ell)}\,,\qquad R=\frac{\cosh^2(\pi \ell)}{\cosh^2(\pi k)+\sinh^2(\pi\ell)}\,.
\end{array}
\end{equation}
Again, $T+R=1$ for $k \in \mathbb R$.
This is possibly the most interesting case, as it shows resonance phenomena. Here, we are assuming that $\ell>0$. Then, both series of pole 
solutions can be written as:
\begin{equation}\label{23}
k_1(n)=\ell-i\left( n+\frac 12\right)\,,\qquad k_2(n)=-\ell-i \left( n+\frac 12\right)\,,
\end{equation}
where $n=0,1,2,\dots$\,. For each value of $n$, solutions $k_1(n)$ and $k_2(n)$ give a pair of resonance poles. Note that, as expected, they are located in the lower half of the $k$ plane symmetrically with respect to the imaginary axis. Let us write each pair of resonance poles as $k_1(n)=\ell-i\gamma_n$ and 
$k_2(n)=-\ell-i\gamma_n$ with $\gamma_n=n+1/2$. Then, the corresponding energy levels are:
\begin{equation}\label{24}
z_R=\dfrac{\hbar^2}{2m}\,k_1(n)^2=E_R- i\,\dfrac{\Gamma}2\,,\qquad
z_R^*=\dfrac{\hbar^2}{2m}\,k_2(n)^2=E_R+i\,\dfrac{\Gamma}2\,
\end{equation}
with
\begin{equation}\label{25}
E_R= \frac{\hbar^2}{2m}\, \left(\ell^2 -  \gamma_n^2 \right) \,, 
\qquad   
\Gamma= \frac{\hbar^2}{2m}\, 4\ell \gamma_n\,.
\end{equation}
\begin{figure}
\centering
\includegraphics[width=0.4\textwidth]{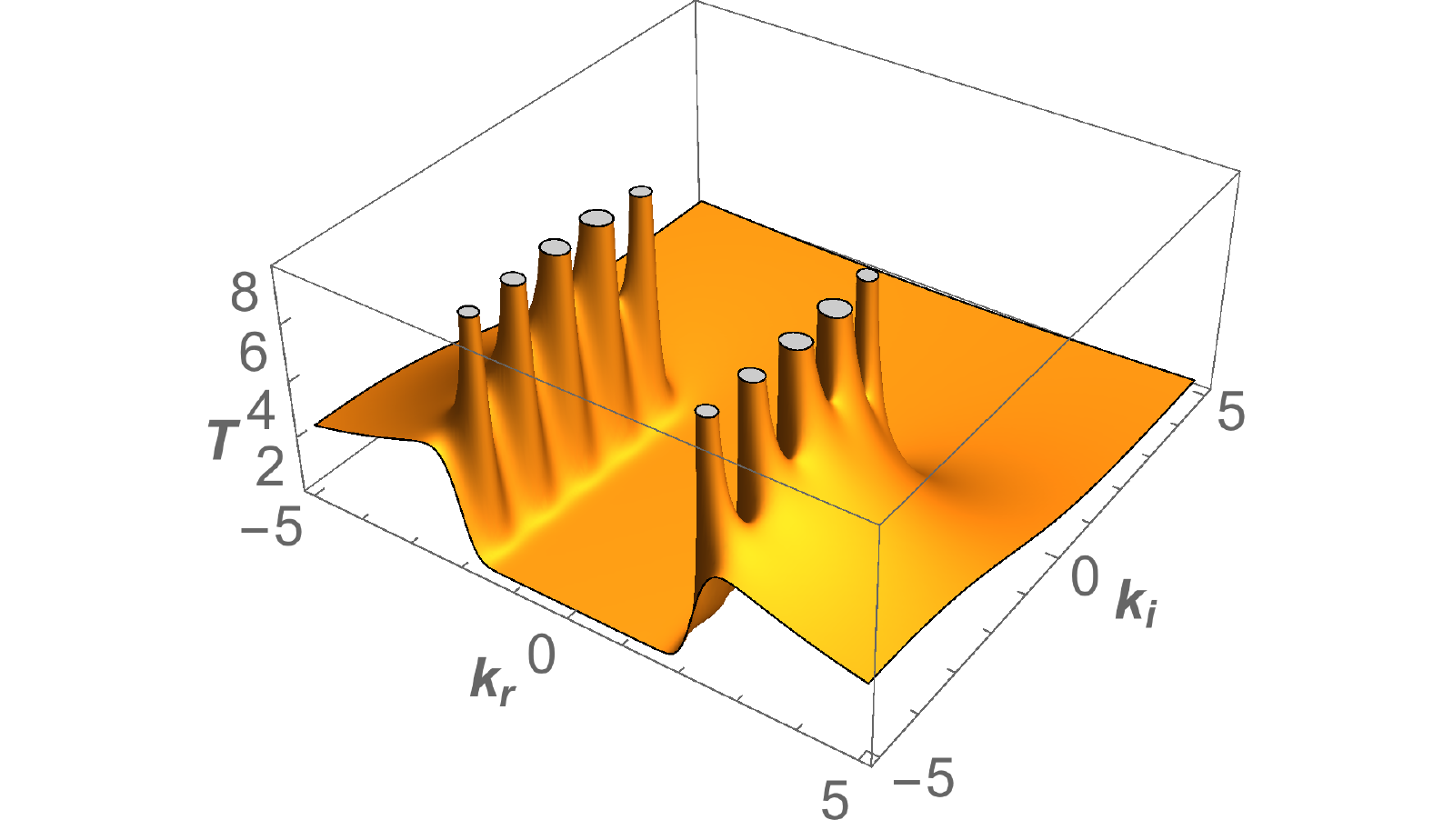}
\ 
\includegraphics[width=0.4\textwidth]{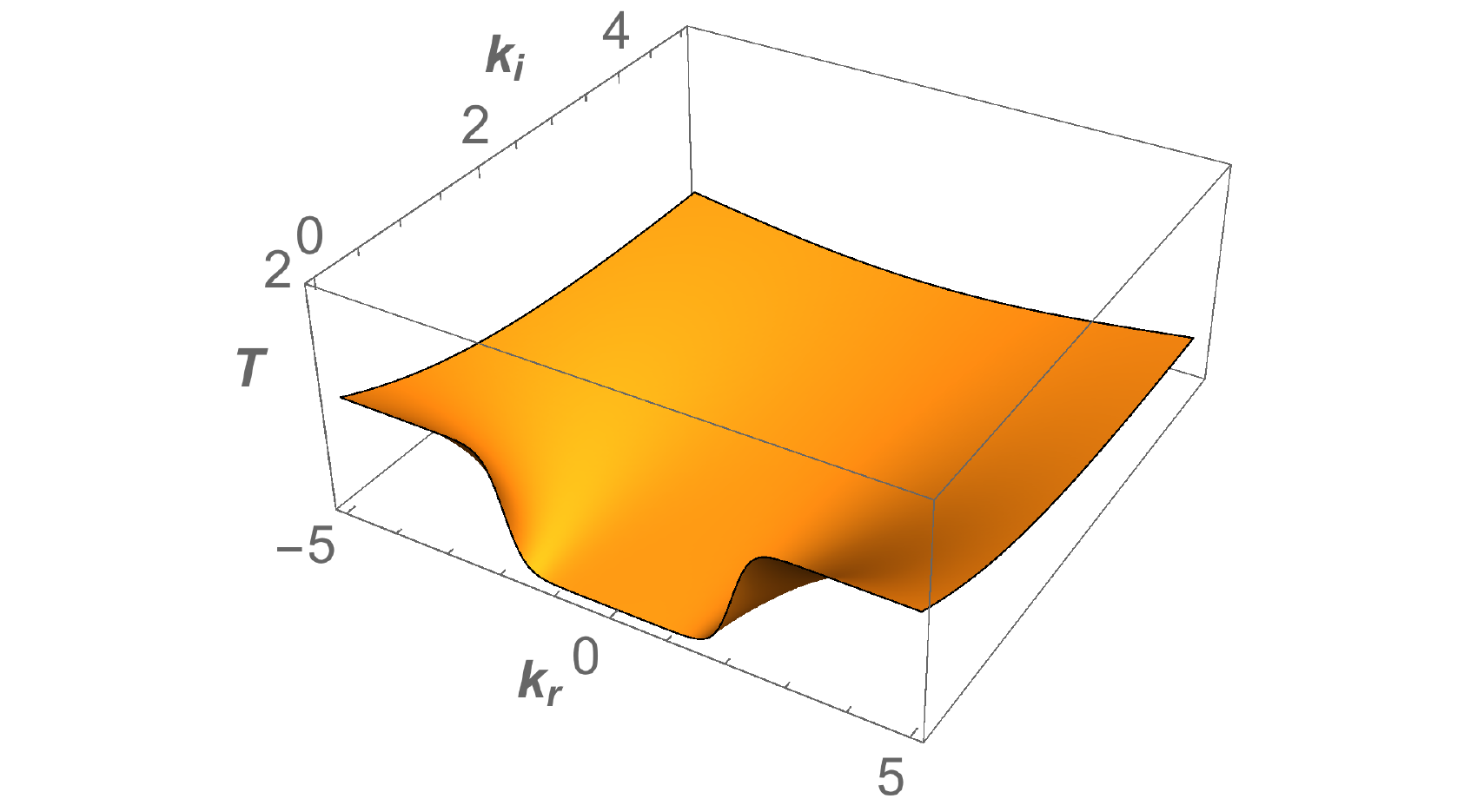}
\caption{\small High barrier potential: Plot of $T(k)$ for  
$\lambda=1/2+ i 2$ and for complex values $k = k_r + i k_i$. 
The singularities are shown at $k_2(n)$:  
$\pm 0.5 - i\, 0.5$,
  $\pm 0.5 - i\, 1.5$,
  $\pm 0.5 - i\, 2.5$,  $\pm 0.5 - i\, 3.5$,  $\pm 0.5 - i\, 4.5$ (left).
  At the right, it is shown the shape of
$T(k)$ when $k_i=0$. This coincides with the value of the transmission coefficient.
}
\label{polos_high}
\end{figure}

\begin{figure}
\centering
\includegraphics[width=0.4\textwidth]{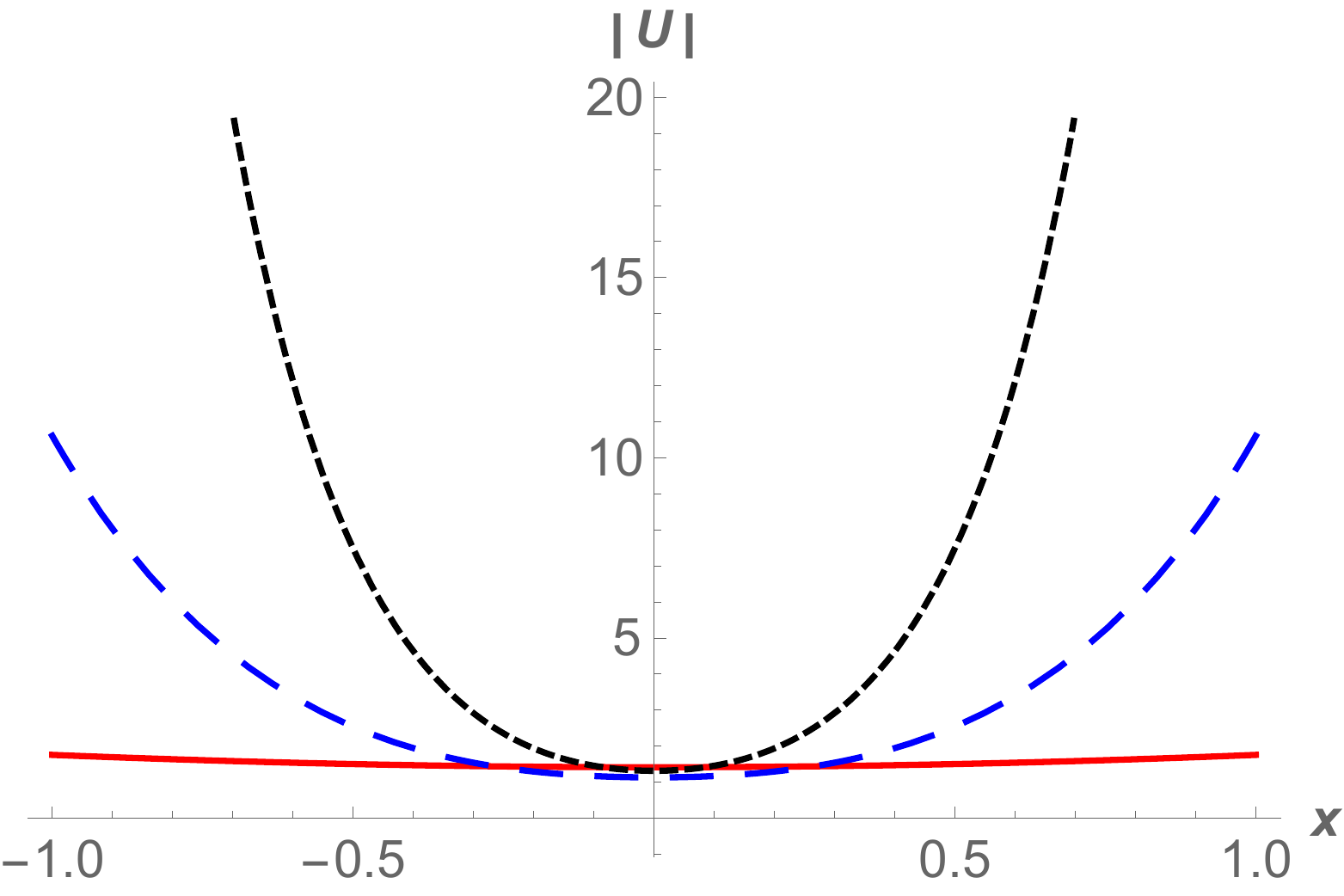}\qquad
\includegraphics[width=0.4\textwidth]{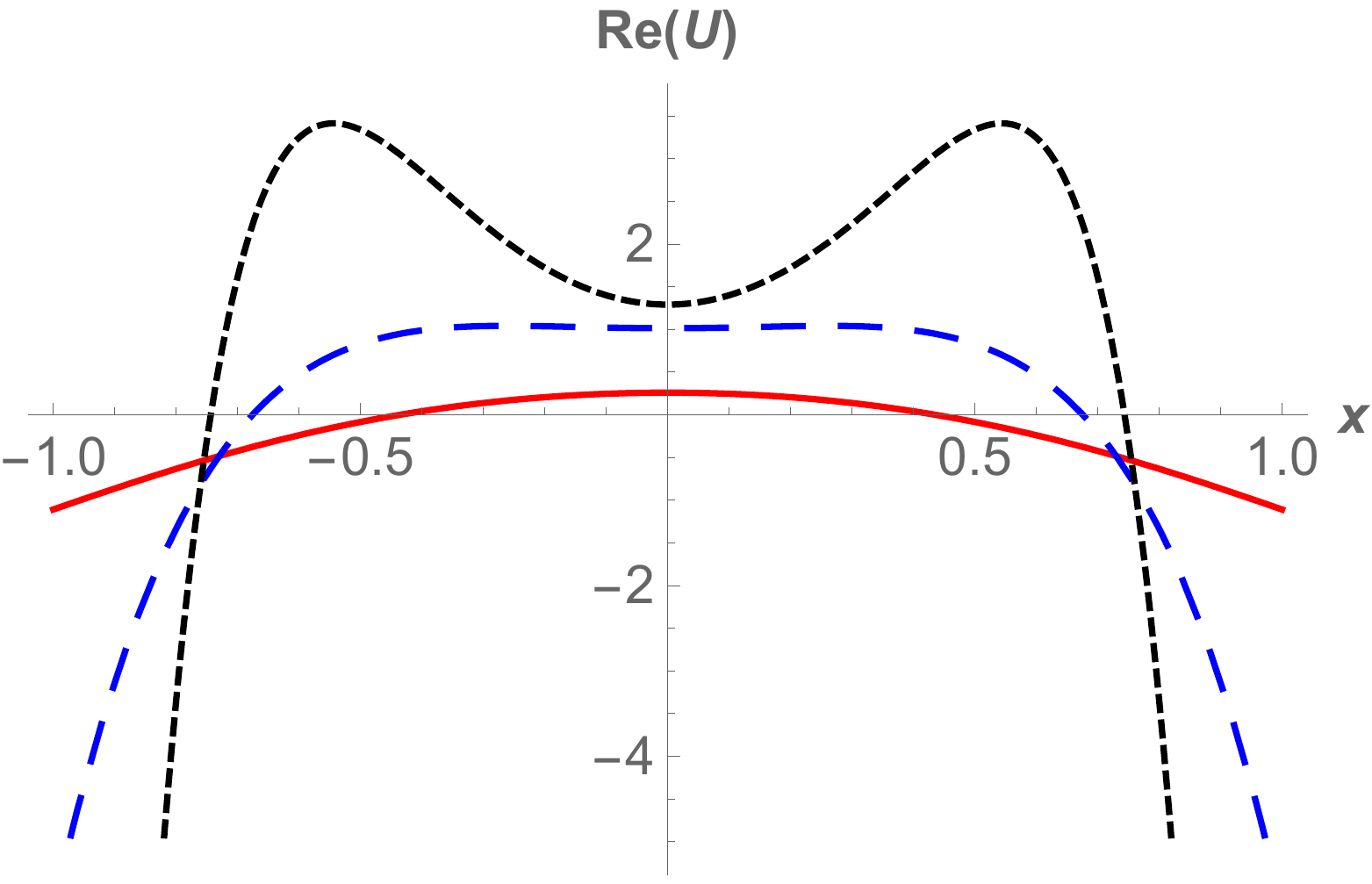}
\caption{\small High barrier potential: Plot of the absolute values of the wave functions for (even) resonances corresponding to $\lambda=1/2+i\, 2$ and resonance poles at $k_1(0)=2-i/2$ (continuous line), $k_1(2)=2-i\, 5/2$ (dashed line) and $k_1(4)=2-i\, 9/2$ (dotted line) (left). At the right, it is shown the real part of the same wave functions.}\label{F8}
\end{figure}

As seen on these formulas, both real and imaginary parts of resonances depend on $n$. There do not exist any other singularities of $S(k)$ like bound or antibound poles, see Fig.~\ref{polos_high}.  
 In Fig.~\ref{F8}  we plot the modulus and the real part of  wave functions 
of the poles $k_1(n)$ for the first three even values of $n$. 
For the odd values of $n$ 
the wave functions are odd, so they include zeros in the origin.
We should recall that, as a consequence of the general theory, the modulus of these wave functions are exponentially growing at the infinity \cite{MBG}.
\section{Ladder operators and singularities of the $S$ matrix}

In a previous paper \cite{KN}, we have constructed the ladder operators for the bound states corresponding to the hyperbolic P\"oschl-Teller potential with $\lambda>1$. Their explicit form is given by:
\begin{equation}\label{26}
\begin{array}{l}
\displaystyle B_n^-= -\cosh x\,\partial_x-\sqrt{-E(n)}\,\sinh x\,,\\[2.75ex] B_{n+1}^+= \cosh x\,\partial_x-\sqrt{-E(n)}\,\sinh x\,,
\end{array}
\end{equation}
where $\partial_x$ stands for derivative with respect to $x$ and we have used $\hbar^2/2m=1$ and $\alpha=1$ for simplicity.  The sequence $E(n)$ represents the energies of bound states, which are given by $E(n)=-(\lambda-n-1)^2$, with $n=0,1,..., [\lambda-1]$, $\lambda>1$. Here $[a]$ is for the highest integer
less than $a$. In \cite{KN}, we have also studied the action of the ladder operators on the eigenfunctions $\psi_n(x)$ corresponding to the eigenvalues $E(n)$:
\begin{equation}\label{27}
B_n^- :\psi_n(x) \longmapsto\psi_{n-1}(x),\qquad B_{n}^+ :\psi_{n -1}(x)\longmapsto\psi_{n}(x)\,.
\end{equation}

The question that now arises is if we could construct ladder operators that behave in a similar manner on the antibound states or even on the resonance states (Gamow states), as stated in the Introduction. Thus, our  next goal is to show that similar ladder operators can be effectively constructed for antibound and resonance states.  As previously done, we shall present this study case by case. 
\subsection{Cases $\lambda>1$ and $1/2\leq \lambda<1$}

In equation (\ref{20}), we have given the two sequences of poles: $k_2(n)$, that contains bound and the antibound states  and $k_1(n)$, which contains antibound states only. Their corresponding negative energies are
\begin{eqnarray}
E_1(n)=k_1^2(n)=-(\lambda+n)^2\,, \label{28}
\\[2.Ex]
E_2(n)=k_2^2(n)=-(n-\lambda+1)^2\,, \label{29} 
\end{eqnarray}
with $n=0,1,2,\dots$ .  In order to find the ladder operators, we extend the
formula (\ref{26}), simply replacing $\sqrt{-E(n)}$ by $k_j(n)$, $j=1,2$. This ansatz will be applied to all the cases. Thus, the explicit form of the ladder operators should be in all cases:
\begin{equation}\label{30}
\begin{array}{l}
B_{j,n}^-= -\cosh x\,\partial_x+i\,k_j(n)\,\sinh x \,,
\\[2.Ex]
B_{j,n+1}^+=
\cosh x\,\partial_x+i\,k_j(n)\,\sinh x\, ,
\end{array}
\end{equation}
where the index $j$ refers to each sequence of poles.
Once we have obtained the wave functions that correspond to the antibound states, 
$\varphi_{j,n}(x)$, we have to show that the behavior of the operators (\ref{30}) on 
$\varphi_{j,n}(x)$ just looks like formula (\ref{27}), i.e., 
\begin{equation}\label{31}
B_{j,n}^- :\varphi_{j,n}(x)\longmapsto\varphi_{j,n-1}(x)\,,\qquad
B_{j,n}^+:\varphi_{j,n-1}(x)\longmapsto\varphi_{j,n}(x)\,,
\end{equation}
where $n=0,1,2\dots$\,. In the next subsections we shall check that this
action is indeed valid in all cases.

By means of these operators we can form an algebra. To this end, we introduce the diagonal operator $B_{j,n}^0$, which is defined in terms of its action on the wave functions $\varphi_{j,n}(x)$ as:
\begin{equation}\label{32}
\begin{array}{l}
B_{1,n}^0\,\varphi_{1,n}(x)=- i\,k_1(n)\,\varphi_{1,n}(x)=-(\lambda+n)\,\varphi_{1,n}(x)\,,
\\[2.ex]
B_{2,n}^0\,\varphi_{2,n}(x)=- i\,k_2(n)\,\varphi_{2,n}(x)=-(-\lambda+n+1)\,\varphi_{2,n}(x)\,.
\end{array}
\end{equation}
With this definition,  $B_{j,n}^0$ is diagonal on the vector space spanned by the $\{\varphi_{j,n}(x)\}$. 
Then, the index free operators $B_{j}^0$ and $B_{j}^\pm$ (for each $j$) close a representation of the ${\rm su}(1,1)$ algebra \cite{KN}:
\begin{equation}\label{33}
[B_{j}^0,B_{j}^{\pm}]=\mp B_{j}^{\pm},\qquad[B_{j}^-,B_{j}^+]=B_{j}^{0}\,.
\end{equation}

Depending on the value of $\lambda$, there are different cases of ladder operators:
\begin{itemize}
\item $\lambda$ is positive half-odd integer.
\item $\lambda$ is integer.
\item $\lambda$ is neither integer nor half-odd integer.
\end{itemize}
Now, we study these three cases separately.
\begin{figure}
\centering
\includegraphics[width=0.25\textwidth]{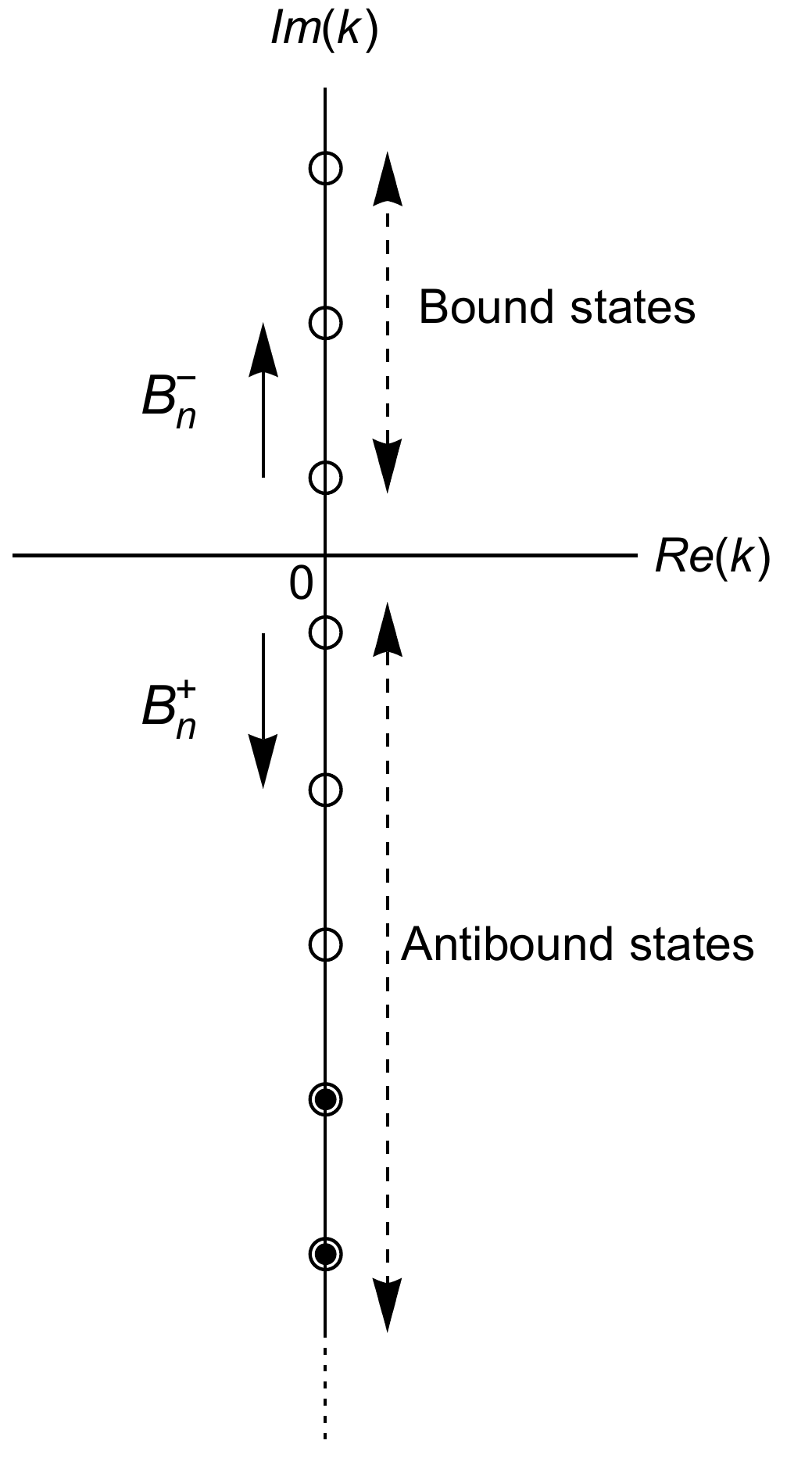}\qquad
\includegraphics[width=0.25\textwidth]{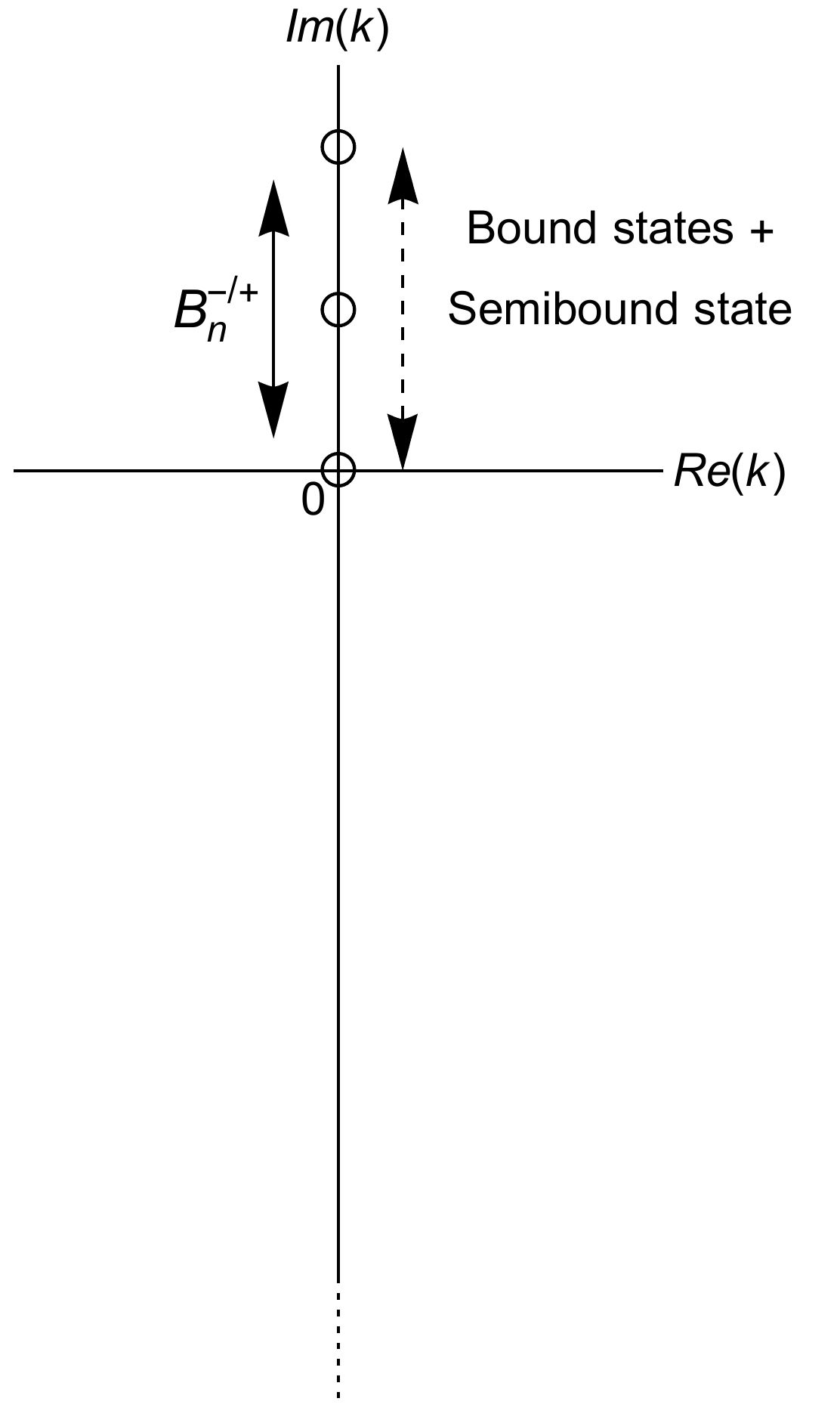}\qquad
\includegraphics[width=0.25\textwidth]{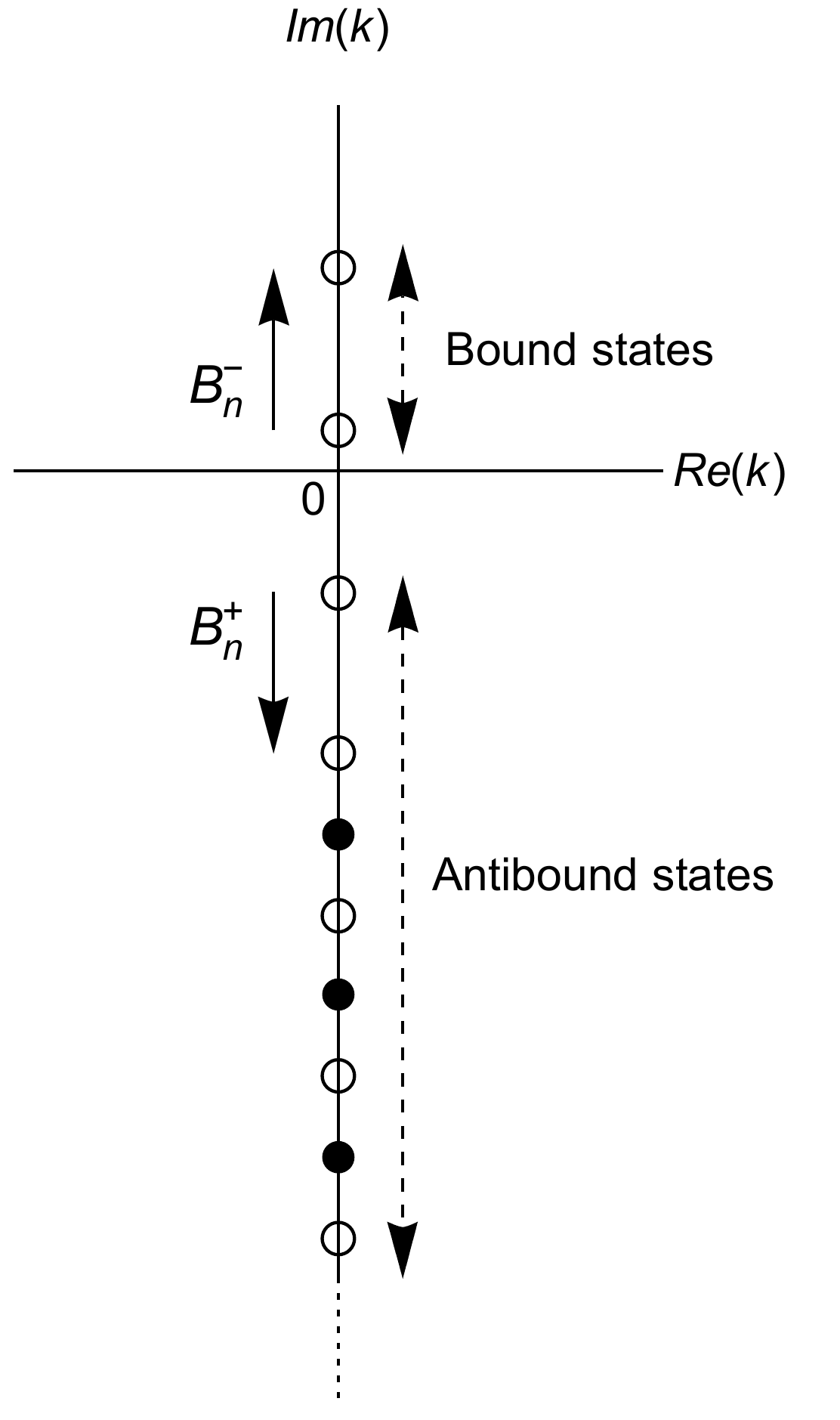}
\caption{\small Action of the ladder operators on the states corresponding to the poles of the series $k_1(n)$ (black disks) and $k_2(n)$ (white disks) for $\lambda=3.5$ (left), $\lambda=3$ (center) and $\lambda=2.25$ (right).  }\label{diagram}
\end{figure}

\subsubsection{$\lambda$ is a positive half-odd integer}

When $\lambda$ is a half-odd positive integer, equations (\ref{20}) or alternatively, equations (\ref{28}) and (\ref{29}) show us that the set of values given by $k_1(n)$ is included in the set of values given by $k_2(n)$, as we can see from this simple formula:
\begin{equation}\label{34}
k_1(n)=k_2(2\lambda-1+n)\,, \qquad n=0,1,2,\dots\,.
\end{equation}
Therefore, we can restrict to the sequence $k_2(n)$.

The number of bound states is given by $[\lambda]$, which is the integer part of $\lambda$ and in this case it coincides with $\lambda-\frac12$.  For instance, if $\lambda=3/2$, we have a unique bound state, if $\lambda=5/2$, we have two and so on. Since the energy of bound states is negative, the ground state corresponds to the highest pole on the imaginary axis, which lies at $k_2(0)=i(\lambda-1)$. This ground state, $\varphi_{2,0}(x)$, is obtained through the condition $B_{2,0}^-\,\varphi_{2,0}(x)=0$, 
\begin{equation}\label{35}
  [-\cosh x\,\partial_x-(\lambda-1)\sinh x]\,\varphi_{2,0}(x)=0\,.
\end{equation}
The solution of (\ref{35}) is quite simple and, as expected, square integrable:
\begin{equation}\label{36}
\varphi_{2,0}(x)=N_0(\cosh x)^{1-\lambda}\,
\end{equation}
where $N_0$ is a normalization constant. Note that $\varphi_{2,0}(x)$ is an eigenfunction of the Hamiltonian $H$ with $\hbar^2/2m=1$ and eigenvalue $E=-(\lambda-1)^2$ as should be. 

In order to obtain the wave functions corresponding to all other bound and antibound states, all we need is to apply successively the creation operators $B_{2,n}^+$. Let us order the poles on the imaginary axis starting with the highest, corresponding to the ground state, and going downwards. Assume the same ordering for their corresponding wave functions. Then, the wave function for the $n$-th pole is given by
\begin{equation}\label{37}
\varphi_{2,n}(x)=N_nB_{2,n}^+B_{2,n-1}^+\dots B_{2,1}^+\,\varphi_{2,0}(x)\,.
\end{equation} 
It is straightforward to check that, indeed, the wave functions obtained 
in this way by $B^+$ coincide with the outgoing wave functions characterized
in (\ref{outgoing}).

Now, we have one stair of ladder operators that connect bound and antibound states as if they were of the same nature. The general form of $\varphi_{2,n}(x)$ is given by
\begin{equation}\label{38}
\varphi_{2,n}(x)=P_n(\sinh x)\,\varphi_{2,0}(x)\,,
\end{equation}
where $P_n(\sinh x)$ is a polynomial of degree $n$ in $\sinh x$. This set of states span a space  that supports the algebra  given by (\ref{33}). It is clear that for some values of $n$, in particular for $0\le n<[\lambda]$, where $[\lambda]$ is the integer part of $\lambda$, the function $\varphi_{2,n}(x)$ is square integrable. The wave functions $\varphi_{2,n}(x)$ with $n\ge [\lambda]$ are not square integrable and are the wave functions for the antibound states. These antibound states include those in the class $k_1(n)$ as in (\ref{34}). 
In Fig.~\ref{diagram} (left), we have chosen $\lambda=3.5$ as an example. There is a unique series of poles and their respective states (wave functions) are related via a unique series of ladder operators. 

Note that for $\lambda=1/2$ and only in this case, $k_1(n)=k_2(n)$ for all values of $n$ as we can see from (\ref{34}). Here there is no bound states but an infinite number of antibound states.

\subsubsection{$\lambda$ is an integer}

This is a very special case, since it corresponds to reflectionless potentials.
When  $\lambda$ is an integer,  the number of bound state poles is given by $\lambda-1$ as can be seen from equation (\ref{29}).  The highest bound state pole is determined by $k_2(0)=i(\lambda-1)$ and the lowest by $k_2(\lambda-2)=i$. 

However, in this case, due to the explicit form of $T_{22}$, we have the following
results.
\begin{itemize}
\item[(i)]
For the values $k_2(n)$, from $n=0$ up to $n=\lambda-2$, the $S$ matrix has singularities,
corresponding to bound states mentioned above.

\item[(ii)]
The value of $k_2(n)$, for $n=\lambda-1$, which gives $k_2=0$, is neither a singularity
nor a zero of the $S$ matrix.

\item[(iii)]
There is no other singularity for $k_1(n)$ or $k_2(n)$.
For the values $k_2(n)$, from $n=\lambda$ up to $n=2\lambda-1$, the $S$ matrix 
instead of singularities, it has zeros.
\end{itemize}

\begin{figure}
\centering
\includegraphics[width=0.4\textwidth]{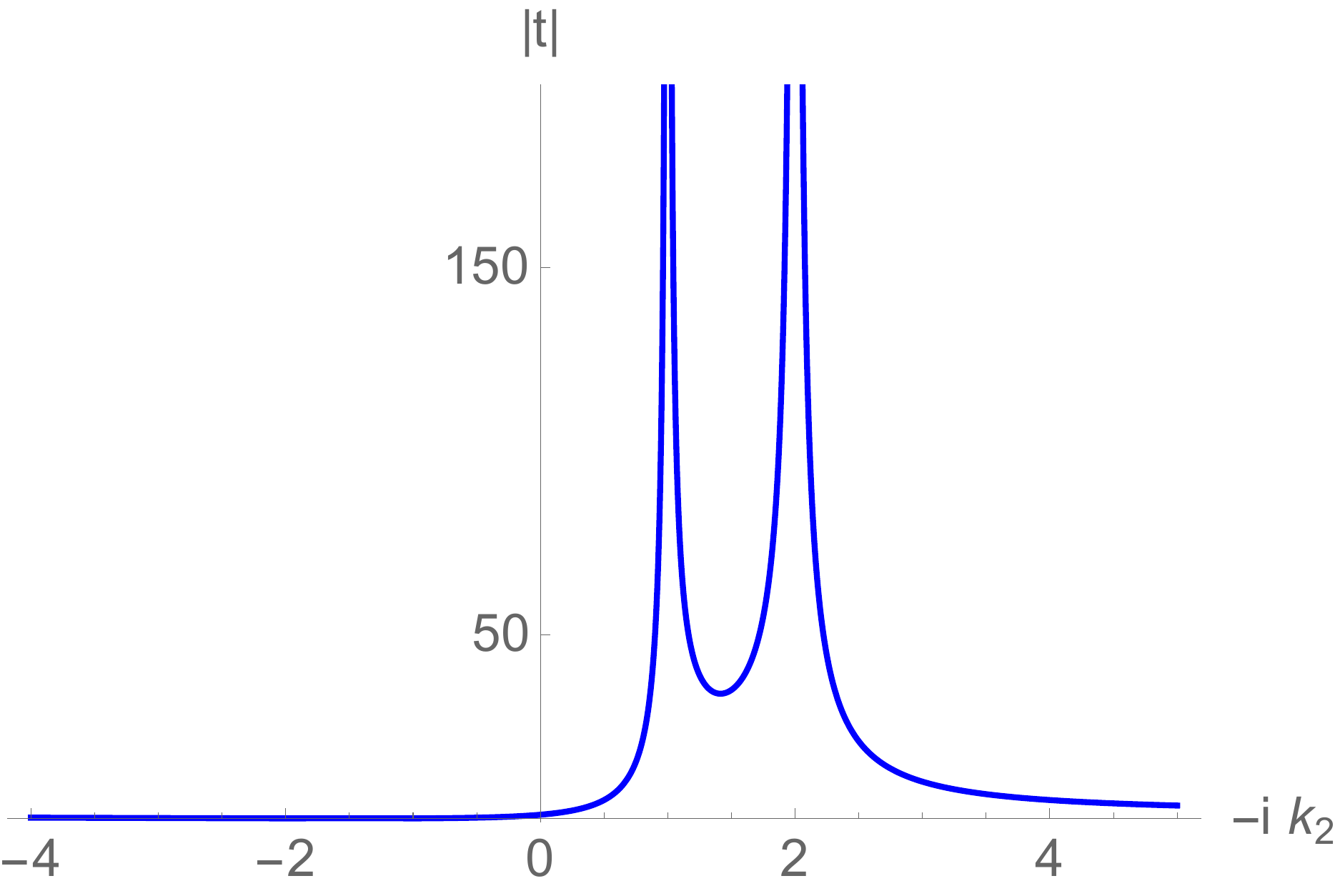}\qquad
\includegraphics[width=0.4\textwidth]{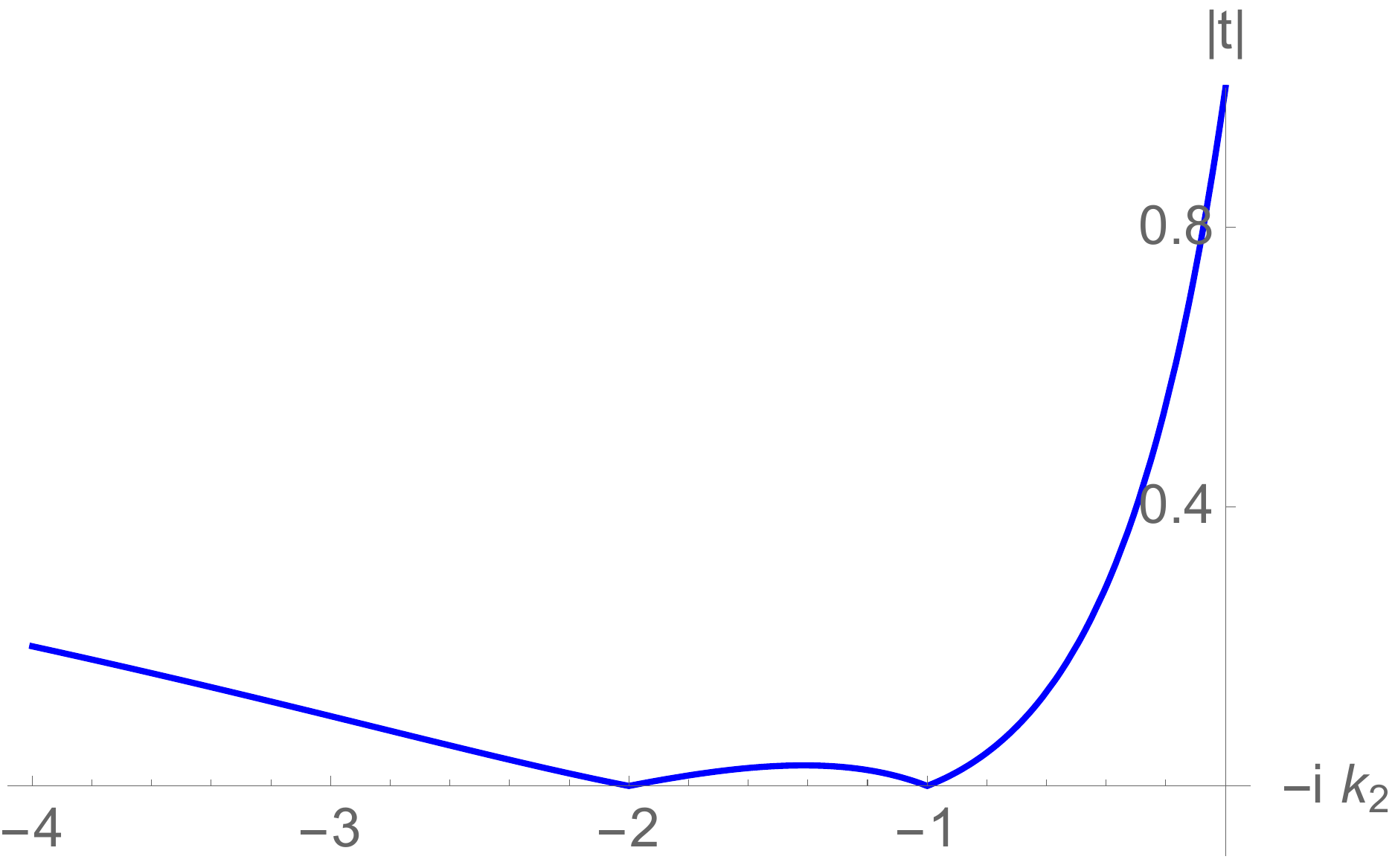}
\caption{\small Plot of $|t| = |1/T_{22}|$ for $\lambda=3$ for the 
imaginary values of $k_2$. At the left, we can appreciate the two singularities
at $k_2=i, i2$ corresponding to bound states. At the right, it is seen  
in more detail the two zeros of this function at $k_2= -i, -i2$\,. }\label{Tentero}
\end{figure}

These properties are illustrated in Fig.~\ref{Tentero} where we represent  $|t| = |1/T_{22}|$,
for $\lambda=3$.  In this case, the ladder operators connect the bound states with the state corresponding 
to $k=0$, which is neither a bound nor an antibound state. 
The central diagram of Fig.~\ref{diagram} corresponds to this case for  $\lambda=3$. In summary, for the reflectionless potentials there are no antibound and resonance poles.

\subsubsection{$\lambda$ is neither a half-odd integer nor an integer}

In this situation, the two sequences given by $k_j(n)$, $j=1,2$ do not coincide. Each of the sequences is connected by its own sequences of ladder operators. The explicit construction does not differ from the previous cases. For example, in Fig.~\ref{diagram} (right), we have chosen $\lambda=2.25$. In this case, we have two different sequences of poles on the imaginary axis with respective different and not connected sequences of ladder operators. 
\subsection{Ladder operators for resonance states}

As we have seen, resonance poles appear  when $\lambda=i\ell+\frac12$ only. Then, the $S$ matrix has an infinite number of resonance poles which appear in pairs symmetrically located with respect to the negative part of the imaginary axis (as always, we have chosen $\alpha=1$):
\begin{equation}\label{41}
k_1(n)=\ell-i\left(n+\frac{1}{2}\right)\,, \qquad   k_2(n)=-\ell-i\left(n+\frac{1}{2}\right)\,,
\end{equation}
where $n=0,1,2,\dots$\,.  
Then, according to (\ref{30}), there is one set of ladder operators for $k_1(n)$ and another one for $k_2(n)$. Both sets are independent. The ladder operators for $k_1(n)$ are given by:
\begin{equation}\label{42}
\begin{array}{l}
B_n^-=-\cosh x\,\partial_x+(i\ell+n+\frac12)\sinh x\,,\\[2.Ex] B_{n+1}^+= \cosh x\,\partial_x+(i\ell+n+\frac12)\sinh x\,,
\end{array}
\end{equation}
with $n=0,1,2,\dots$. The sequences for $k_2(n)$ are obtained just by replacing $i\ell$ by $-i\ell$ in (\ref{42}). These two types of operators act as in (\ref{31}) on the decaying and growing Gamow states, respectively. Each sequences of operators satisfy commutation relations as in (\ref{33}).

We obtain the first decaying  Gamow state, corresponding to $n=0$, by solving the differential equation  $B_0^-\varphi_0(x)=0$ with $B_0^-$ as in (\ref{42}). For the first growing Gamow state, we just replace $\ell$ by $-\ell$ in (\ref{42}) with $n=0$. Up to a constant factor, we obtain
\begin{equation}\label{43}
\!\!\!\varphi_0^D(x)= (\cosh x)^{i\ell+1/2}\,,\qquad \varphi_0^G(x)= (\cosh x)^{-i\ell+1/2}\,,
\end{equation}
where the superscripts $D$ and $G$ stand for decaying and growing. Here,  $i\,k_1(0)={i\ell+1/2}$ and $i\,k_2(0)={-i\ell+1/2}$, respectively. Other Gamow states are found by successive application of creation operators. They have the form $\varphi_n^D(x)=P_n(\sinh x)\varphi_0^D(x)$ and $\varphi_n^G(x)=P_n(\sinh x)\varphi_0^G(x)$, respectively for decaying and growing Gamow states. Here, the polynomials of degree $n$, $P_n(\sinh x)$, have complex coefficients. Note that $H\varphi_n^D(x)=[k_1(n)]^2\varphi_n^D(x)$ and $H\varphi_n^G(x)=[k_2(n)]^2\varphi_n^G(x)$. These Gamow states obtained by
ladder operators coincides with the solutions previously found in
(\ref{outgoing}).

In conclusion, the set of decaying and growing Gamow vectors span two different spaces. Both serve as support of the Lie algebra $su(1,1)$ spanned by the ladder operators. 

\section{Supersymmetric partners using antibound and resonance states}

A supersymmetric   partner of a given Hamiltonian $H=-\partial^2_x+V(x)$ is another Hamiltonian $\widetilde H=-\partial^2_x+\widetilde V(x)$ constructed following a standard recipe.  
Let $E(0)$ be the energy of the ground state of $H$. 
Next, we shall take a 
solution $H\psi(x)=\varepsilon\psi(x)$ of the Schr\"odinger equation, such that 
(i) $\varepsilon<E(0)$, (ii) $\psi(x)$ have no zeros, and (iii)  $1/\psi(x)$ be square integrable. Then, construct the function $W(x):=\psi'(x)/\psi(x)$, where $\psi'(x)=d \psi(x)/dx$. This function $W(x)$  is called the superpotential. Then, define the shift operators $A^\pm$ as:
\begin{equation}\label{44}
A^\pm:=\pm\partial_x+W(x)\,.
\end{equation}
Using the shift operator (\ref{44}), one can show that $H$ can be factorized as \cite{CHIS,BOG1}
\begin{equation}\label{45}
H=A^+A^-+\varepsilon=-\partial^2_x+W^2(x)+W'(x)+\varepsilon=-\partial^2_x+V(x)\,,
\end{equation}
where $W'(x)=dW(x)/dx$. Then, the  supersymmetric partner, $\widetilde H$, of $H$ is obtained by reversing the order of the factor operators in the form
\begin{equation}\label{46}
\widetilde H= A^-A^++\varepsilon= -\partial^2_x+W^2(x)-W'(x)+\varepsilon = -\partial^2_x+\widetilde V(x)\,.
\end{equation}

The potential $\widetilde V(x)$ is also called the supersymmetric partner potential of  $V(x)$. Then, the new Hamiltonian $\widetilde H$ has one more bound state with wave function $\widetilde{\psi}(x)={1}/{\psi(x)}$ and energy $\varepsilon<E(0)$.   However, the other bound states of $\widetilde H$ are obtained from the bound states of $H$ by means of the action of $A^-:\tilde{\psi}_n(x) \propto A^-\psi_n(x)$ for $n=1,2,\dots$. 
\subsection{SUSY partners by antibound states}

The well ($\lambda>1$) and the low barrier ($1/2\leq\lambda<1$) P\"oschl-Teller potentials have anitibound states. If these states are represented by functions with square integrable inverse, these functions can be used  to produce SUSY partners with discrete spectrum. 
\begin{figure}
\centering
\includegraphics[width=0.4\textwidth]{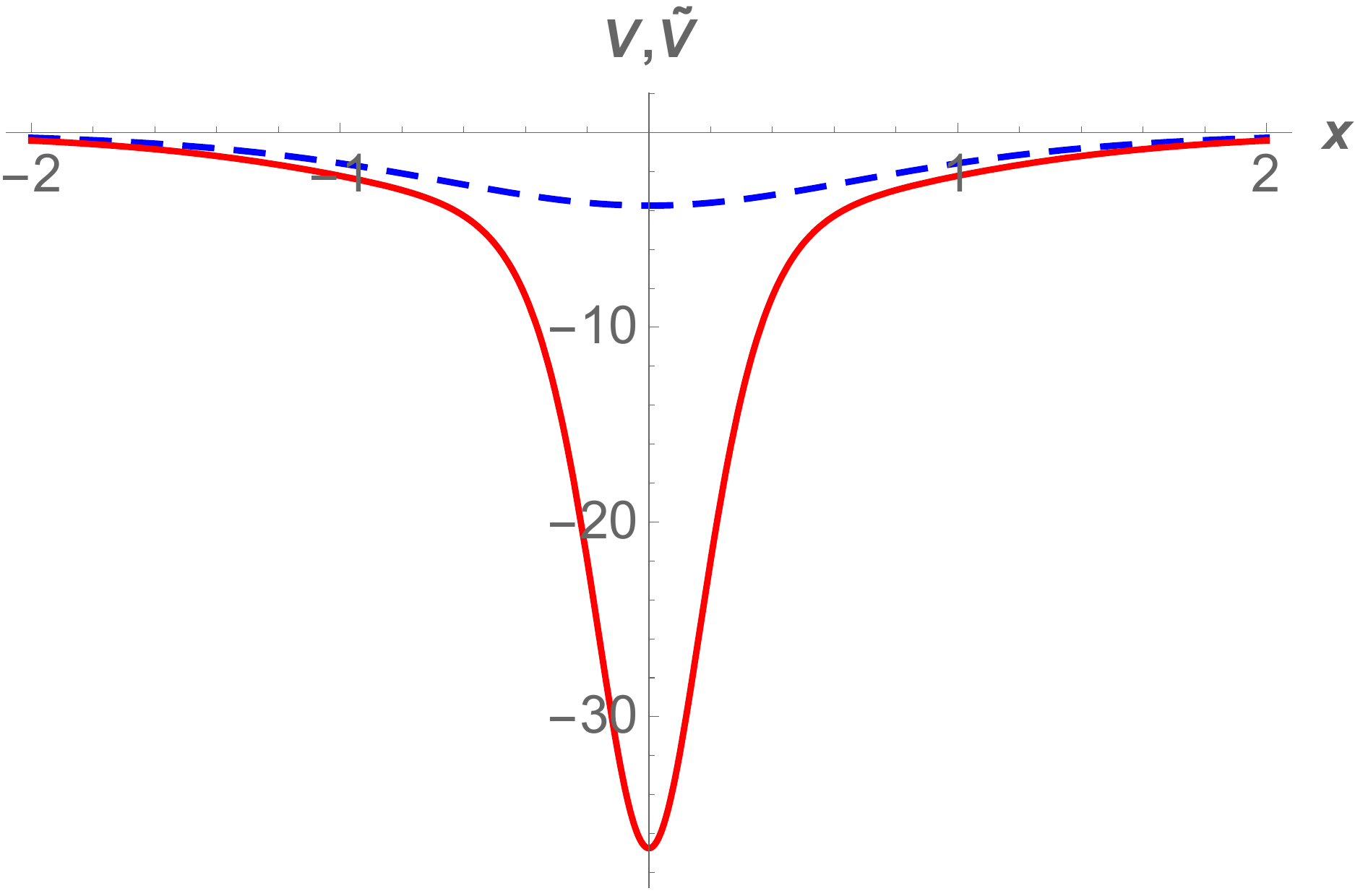}
\caption{\small P\"oschl-Teller well potential with $\lambda=2.5$ (dashed) versus its potential partner  (\ref{48}) (continuous).}\label{fig11}
\end{figure}
As an example, take $\lambda=2.5$. This value corresponds to a potential well. This potential well has two bound states as shown by $k_2(n)$ in (\ref{20}): $E_2(0)=k_2(0)^2=-1.25$, $E_2(1)=k_2(1)^2=-0.25$. The antibound state corresponding to the pole $k_2(6)=-i 4.5$, with energy $\varepsilon=E_2(6)=-20.25$, is positive for all values of $x\in\mathbb R$ and is
\begin{equation}\label{47}
\varphi_{2,6}(x)=(1+7\sinh^2 x)(\cosh x)^{5/2}\,,
\end{equation}
where we have choosen the positive branch for the square root. Hereafter we will use the notation $\varphi_{2,6}(x):=\varphi_{6}(x)$ for simplicity. We use $\varphi_6(x)$ to construct the superpotential  $W(x)=\varphi'_6(x)/\varphi_6(x)$. Then,  the initial and partner potentials  have the following  form:
\begin{equation}\label{48}
\begin{array}{l}
\displaystyle V(x)=-\frac{15/4}{\cosh^2{x}},\\[2.Ex] 
\displaystyle \widetilde V(x)= -\frac{21(-161+55\cosh 2x+120\, {\rm sech}^2 x)}{2(5-7\cosh 2x)^2}\,.
\end{array}
\end{equation}
The general theory \cite{CHIS} shows that the Hamiltonian $\widetilde H=-\partial^2_x+\widetilde V(x)$  has one more bound state than the initial P\"oschl-Teller potential. This bound state has precisely the energy $\varepsilon=E_2(6)=-20.25$. 
In Fig. \ref{fig11}, we plot the P\"oschl-Teller potential well for $\lambda=2.5$ as well as the partner potential $\widetilde V(x)$. We observe that the partner potential is deeper than the original one, so that it seems natural that the latter has one more bound state. 
In Fig.~\ref{fig12}, we plot the wave functions for the antibound state $\varphi_6(x)$ and the new bound state of $\widetilde{H}$, $\widetilde{\varphi}_6(x)=1/{\varphi_6(x)}$.
\begin{figure}
\centering
\includegraphics[width=0.4\textwidth]{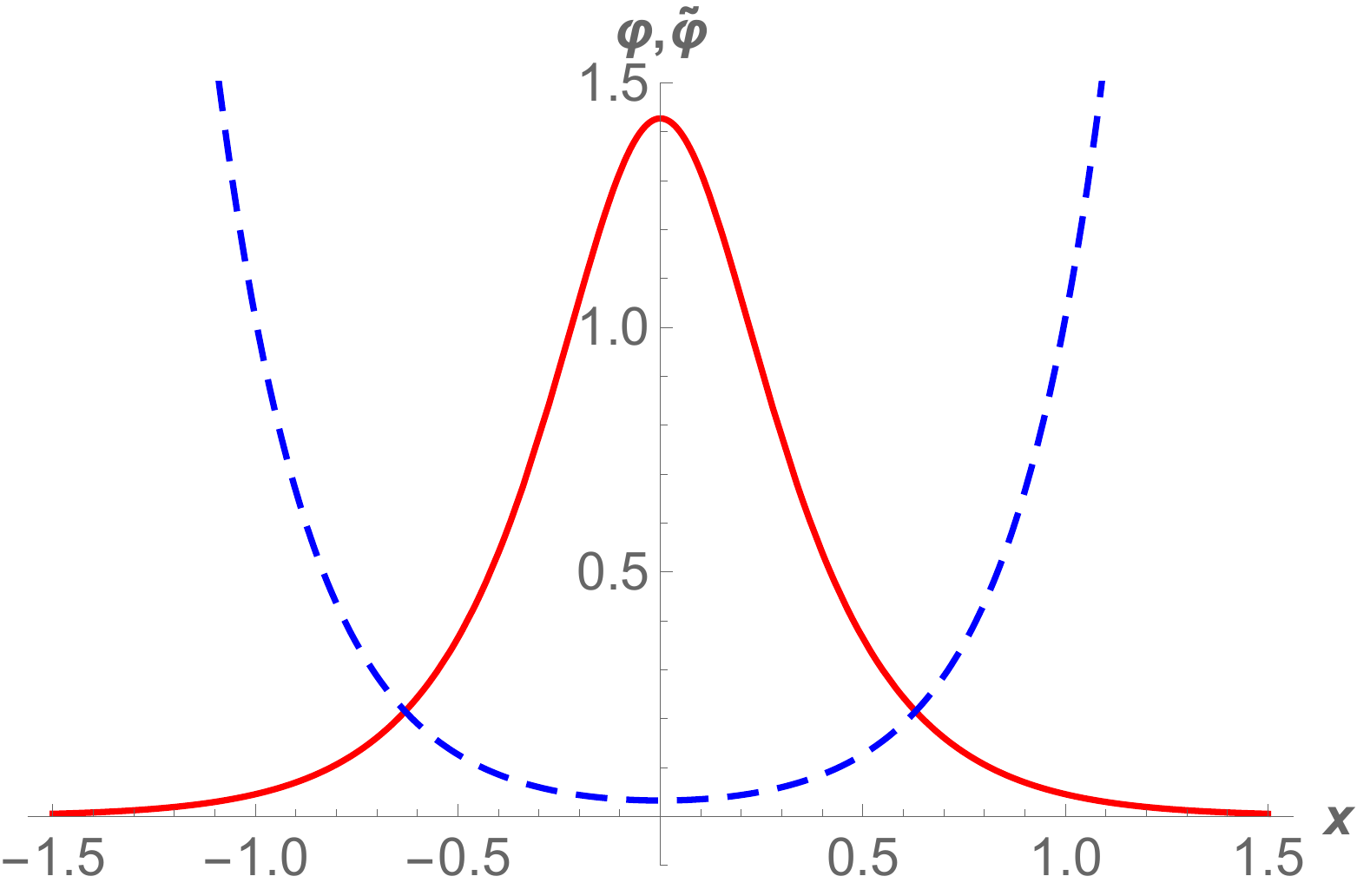}
\caption{\small Plot of the wave function corresponds to antibound state $\varphi_6(x)$ as in (\ref{47})  (dashed) 
and the wave function for bound state $\tilde{\varphi}_6(x)=1/\varphi_6(x)$ (continuous).}\label{fig12}
\end{figure}
\subsection{Complex potentials as SUSY partners of the P\"oschl-Teller potential}

A quite tantalizing possibility is the construction of partners potentials of the P\"oschl-Teller potential, using Gamow states. This will produce series of complex potentials,
$\widetilde{V}=V_{re}+i\, V_{im}$, for which their properties have to be explored. 

Let us consider a high barrier P\"oschl-Teller potential with $\lambda=1/2+i \ell$ and $\ell=3$. One pair of resonance poles in momentum representation are located at $\pm 3-i 5/2$. Let us choose the pole at $k_1(2)=3-i5 /2$, which lies on the forth quadrant. This means that its Gamow state is a decaying Gamow state. One can easily obtain its explicit form from (\ref{37}), (\ref{42}) and (\ref{43}):
\begin{equation}\label{49}
\varphi_2^D(x)= \left(1+(3+i 6)\sinh^2 x\right)(\cosh x)^{i 3+1/2}\,.
\end{equation}

Note that $\varphi_2^D(x)$ never vanishes at any point, so that its inverse $1/ \varphi_2^D(x)$ is always well defined. In addition, it is not difficult to check that $1/ \varphi_2^D(x)$ is square integrable. According to the general theory, the new partner Hamiltonian $\widetilde H=-\partial_x^2+\widetilde V(x)$ has the bound eigenfunction $1/ \varphi_2^D(x)$ with eigenvalue  
$\varepsilon= E_1(2)=(3-i 5/2)^2$. The initial and the partner potential are given by:

\begin{equation}\label{50}
\begin{array}{l}
\displaystyle V(x)=\frac{37/4}{\cosh^2{x}},\\[2.Ex]
\displaystyle \widetilde V(x)=\frac{15\, ((-95+i 236)+(124-i 448)\cosh 2x)-15\, ((37-i 148)\cosh 4x}{ 8\,{\cosh}^2 x\,((1+i 6)-(3+i 6)\cosh 2x)^2}\,.
\end{array}
\end{equation}

\begin{figure}
\centering
\includegraphics[width=0.4\textwidth]{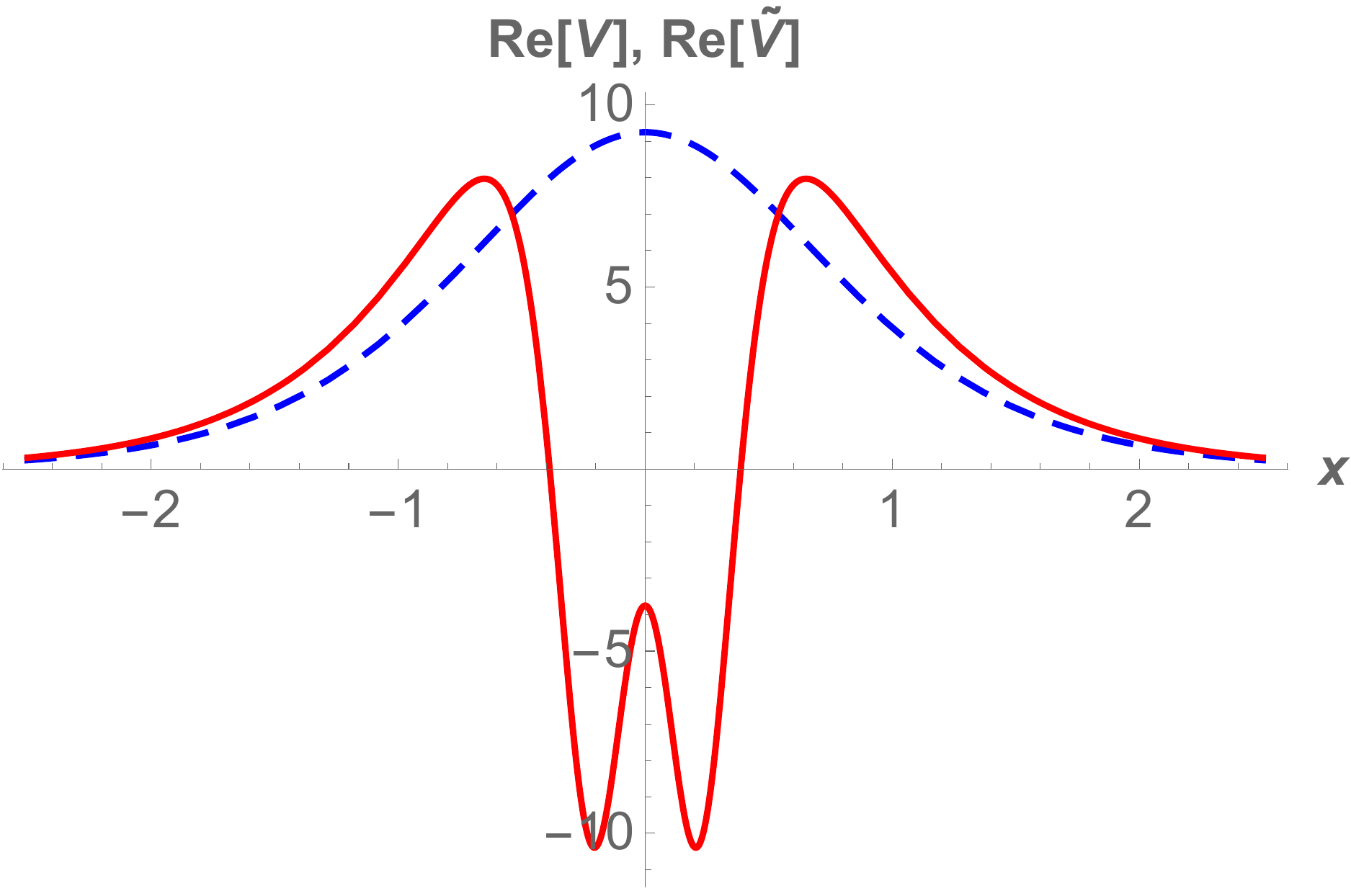}\qquad
\includegraphics[width=0.4\textwidth]{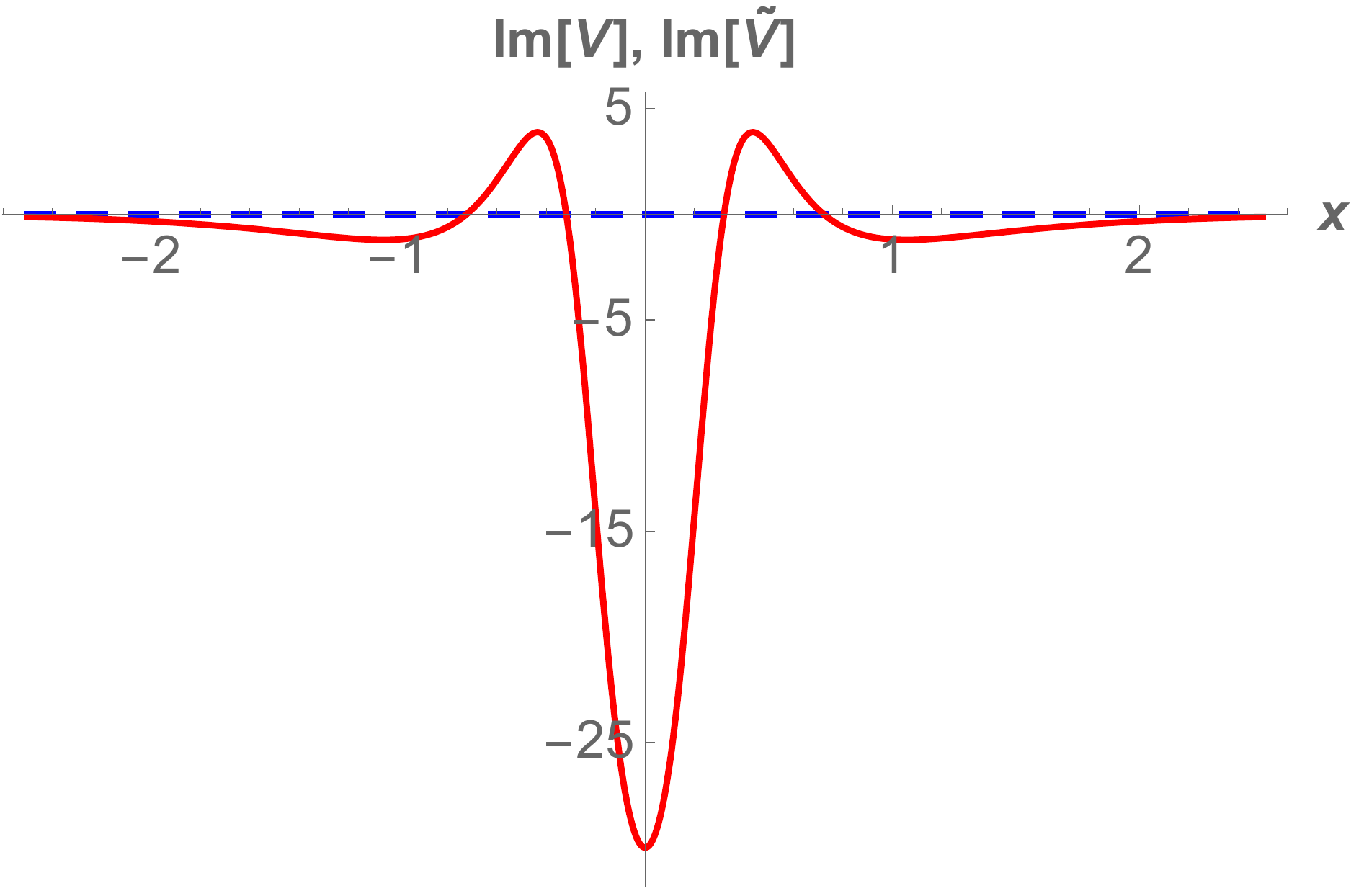}
\caption{\small Real and imaginary parts of the partner potential (\ref{50}) (continuous line) plotted versus the original potential (dashed line).}\label{fig13}
\end{figure}

In Fig.~\ref{fig13}, we plot the partner potentials given by (\ref{50}). Needless to say that the imaginary part of the original potential is identically zero and original potential is real. 
\begin{figure}
\centering
\includegraphics[width=0.4\textwidth]{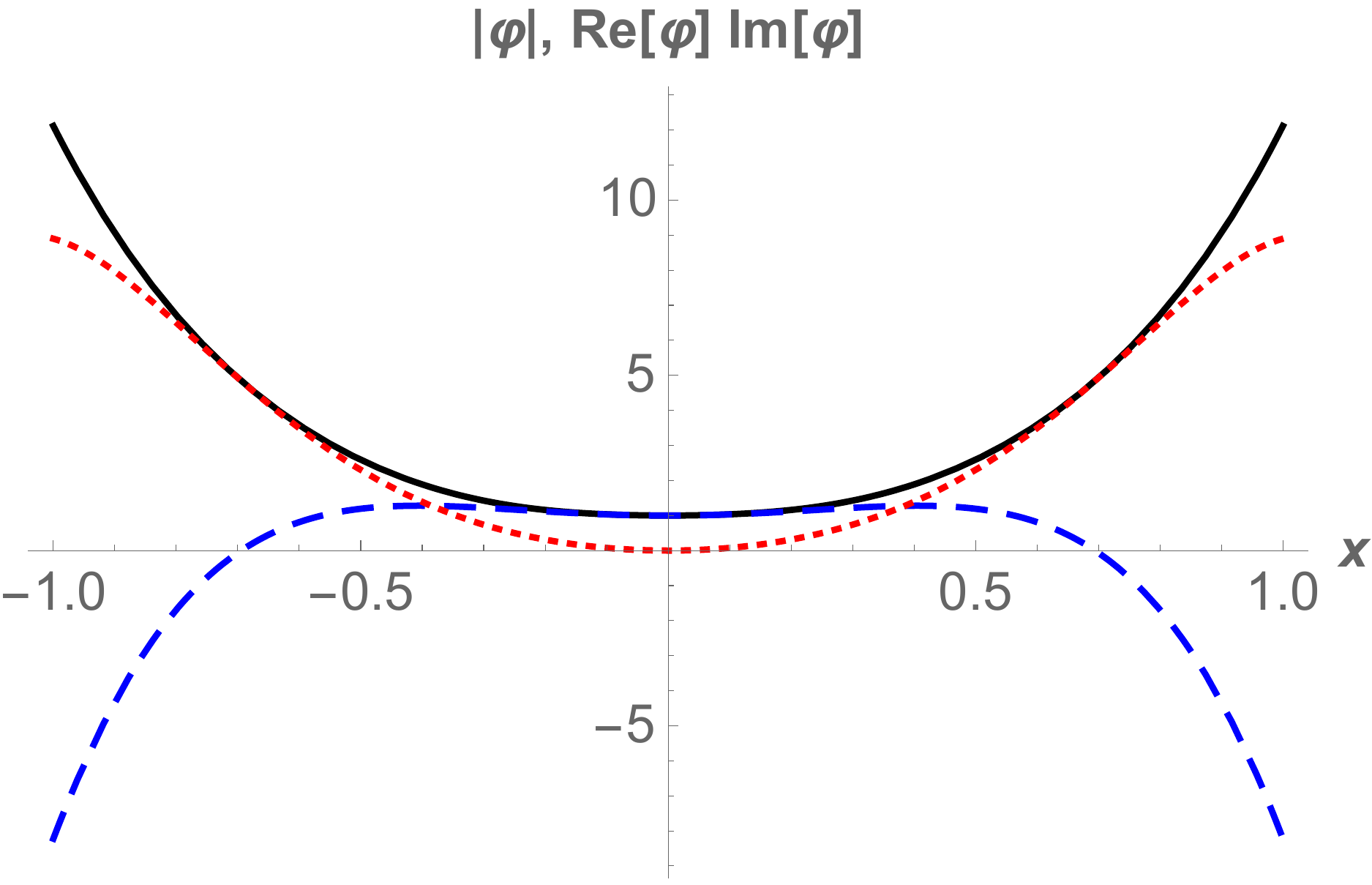}\qquad
\includegraphics[width=0.4\textwidth]{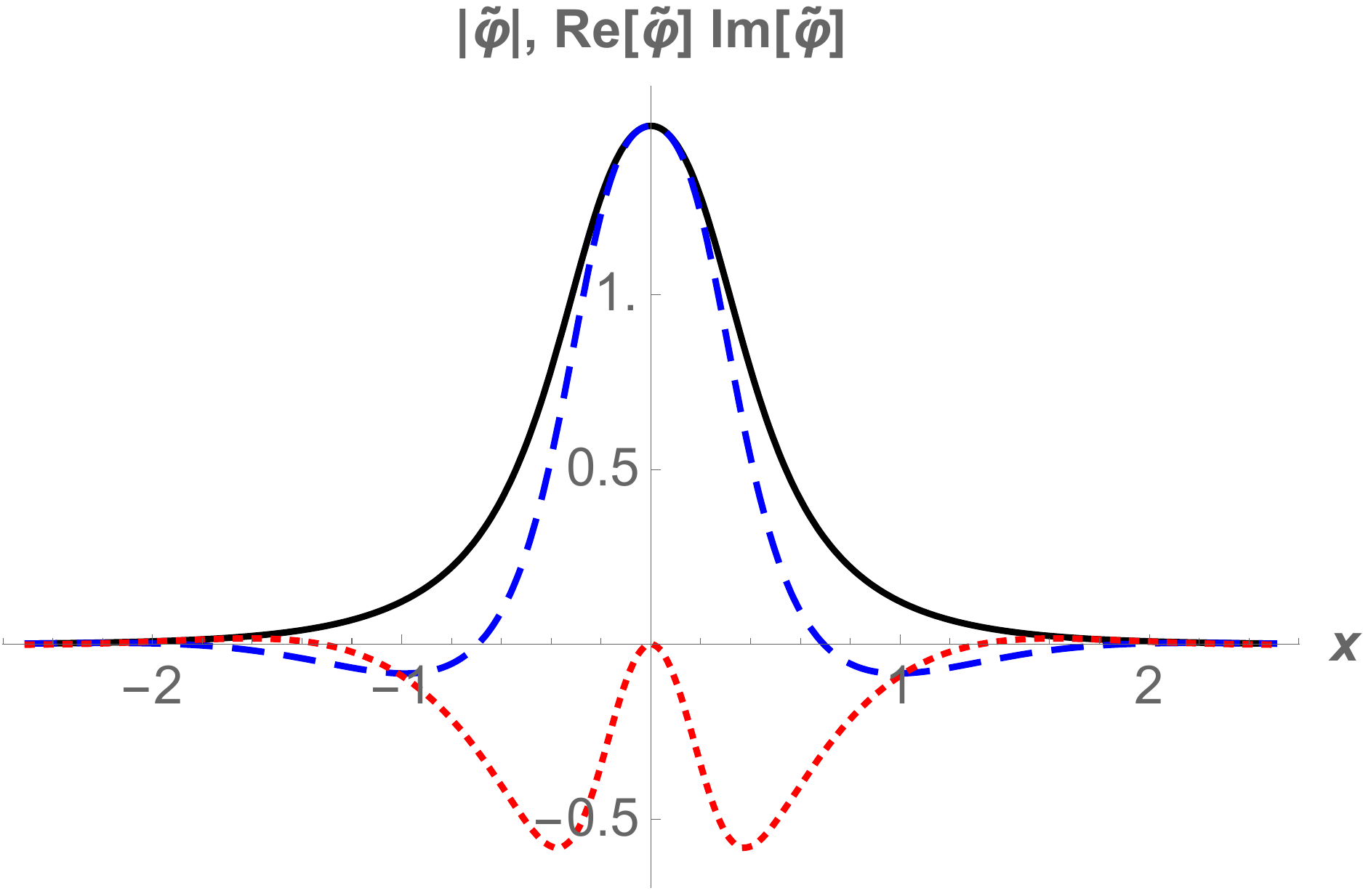}
\caption{\small At the left, it is modulus (continuous line), real Á(dashed line) and imaginary parts (dotted line) of the Gamow state $\varphi_2^D(x)$ in (\ref{49}). At the right, it is the same for the normalizable  wave function $1/\varphi_2^D(x)$.}\label{fig14}
\end{figure}
In Fig. \ref{fig14}, we plot at the left the Gamow state $\varphi_2^D(x)$ given by (\ref{49}) and at the right, the normalizable eigenfunction $1/\varphi_2^D(x)$ with eigenvalue $E_1(2)=(3-i5/2)^2$. 
\section{Concluding remarks}

The P\"oschl-Teller potential has different shapes depending on the parameter $\lambda$ as follows: a well ($\lambda>1$) , a low barrier ($1/2\leq\lambda<1$) or a high barrier ($\lambda=1/2+i\,\ell$).
The properties of the $S$ matrix ($S(k)$) also change depending on the parameter $\lambda$. 
In the momentum representation, the poles of $S(k)$ corresponding to purely outgoing  condition on the wave functions have been determined analytically.
As expected, the well potential is the only one that may have bound states, represented by simple poles of $S(k)$ on the positive imaginary semi-axis. In addition, it has an infinite number of antibound (virtual states), characterized by simple poles of $S(k)$ on the negative imaginary semi-axis. 
There is one important exception for the integers values of $\lambda$, 
corresponding to reflectionless potentials where there are no antibound poles.
No resonance poles are present, which is somehow a surprise. The low barrier potential has infinite antibound poles but neither bound nor resonance poles.  In both cases, poles of $S(k)$ of either type on the imaginary axis can be classified into two independent sequences: one is for $k_1(n)$ and the second for $k_2(n)$. For the potential well only the sequence $k_2(n)$ contains both bound and antibound poles.

The situation given by the high barrier is quite different, as $S(k)$ has now only resonance poles, which appear in pairs on the lower half plane, symmetrically located with respect to the imaginary axis. The number of these pairs is infinite.  In this work the poles of the $S(k)$ and the corresponding bound, antibound and resonance states have been determined analytically. 

The eigenvalues of the P\"oschl-Teller Hamiltonian are given by $E=\frac{\hbar^2}{2m}k^2$. In particular, this formula includes the square of the poles as eigenvalues. Eigenfunctions corresponding to bound states are square integrable, but the corresponding eigenfunctions for antibound and resonance poles are not normalizable. In fact, they diverge exponentially as $x\longmapsto\pm \infty$. 

In previous studies concerning Hamiltonians with discrete spectrum, we have obtained creation and annihilation (ladder) operators which relate bound states of the Hamiltonian \cite{SJ1,SJ2,PRON}. In this paper we have shown that this formalism can be extended to wave functions related to  antibound and resonance poles which have been found for this potential. We have obtained  the explicit forms of these ladder operators for all cases.  Furthermore, we show that these operators satisfy an algebra analogue to the spectrum generating algebra for bound states. This is one of the most interesting result presented in this paper. 
In all cases, there are two sequences of ladder operators corresponding $k_1(n)$ and $k_2(n)$. In the case of the potential well, they connect antibound states for
$k_1(n)$ or bound and antibound states for  $k_2(n)$. In the case of
the low barrier, ladder operators connect antibound states for
both sequences.
Concerning the high barrier, we have two sequences of resonance poles: those in the third and, independently, those in the forth quadrant. Their corresponding eigenfunctions, growing and decaying Gamow states are related by two independent sequence of ladder operators. 

Finally, the factorization method has been applied to this potential in order to obtain new solvable potentials.  Eigenfunctions of the Hamiltonian with real eigenvalues have been used in the literature in order to obtain the so called partners potentials to a given one. We have applied this idea to construct P\"oschl-Teller partner potentials  using antibound eigenfunctions with real negative energies. By means of wave functions of the resonance states
with complex energies we have also obtained complex partner potentials 
having bound states. 
The interest of this kind of complex potentials will be a matter of future work. 

\section*{Acknowledgements}

We acknowledge partial financial support to  the Spanish MINECO (Project MTM2014-57129-C2-1-P). 
D. \c{C}evik and \c{S}. Kuru acknowledge the
warm hospitality at the Universidad de Valladolid, Departamento de F\'isica Te\'orica,  where part of this work has been done. 



\begin{thebibliography}{99}

\bibitem{BOHM} A. Bohm, {\it Quantum Mechanics: Foundations and Applications} (Springer, Berlin and New York, 2001). 

\bibitem{BB} A. Bohm, H.V. Buy, Resonance and decay phenomena lead to quantum
mechanical time asymmetry, J. Phys: Conf. Ser., {\bf 428}, 012016 (2013). 

\bibitem{BK} A. Bohm, P. Kielanowski, Time asymmetric quantum theory
and the Z-boson mass and width, Fort. Phys., {\bf 50}, 496-502 (2002).

\bibitem{BEU} A. Bohm, F. Erman, H. Uncu, Resonance phenomena and time asymmetric quantum mechanics, Turk. J. Phys., {\bf 35}, 209-240 (2011). 

\bibitem{NU} H.M. Nussenzveig, {\it Causality and Dispersion Relations} (Academic, New York and London, 1972).

\bibitem{NAZAR} N. Michel, W. Nazarievich, M. P{\l}oszajczak, T. Vertse, Shall model in the complex energy plane, J. Phys. G: Nucl. Part. Phys., {\bf 36}, 013101 (2009). 

\bibitem{DRSL} J. Darai, A. R\'acz, P. Salamon, R.G. Lovas, Antibound poles in cut-off Woods-Saxon and strictly finite-range potentials, Phys. Rev. C, {\bf 86}, 014314 (2012). 

\bibitem{BG} A. Bohm, M. Gadella, {\it Dirac Kets, Gamow Vectors and Gelfand Triplets}, Springer Lecture Notes in Physics, {bf 348} (Springer, Berlin and New York, 1989). 

\bibitem{CG} O. Civitarese, M. Gadella, Physical and mathematical aspects of Gamow states, Phys. Rep., {\bf 396}, 41-113 (2004). 


\bibitem{AG} I.E. Antoniou, M. Gadella, Irreversibility, resonances and rigged Hilbert spaces, Springer Lecture Notes in Physics, {\bf 622}, 245-302 (2003). 


\bibitem{FGR} L. Fonda, G.C. Girardi, A. Rimini, Decay theory for unstable quantum systems, Rep. Progr. Phys., {\bf 41}, 587-631 (1978).


\bibitem{KKH}  V.I. Kukulin, V.M. Krasnopolsky, J. Horacek, {\it Theory of Resonances. Principles and Applications} (Academia, Prag, 1989).

\bibitem{RSIII} M. Reed, B. Simon, {\it Scattering Theory} (Academic, New York, 1979). 

\bibitem{AJS} W.O. Amrein, J.M. Jauch, K.B. Sinha, {\it Scattering Theory in Quantum Mechanics} (Benjamin, London, 1977). 

\bibitem{YAF} D. Yafaev, {\it Scattering Theory: Some Old and New Problems}, Springer Lecture Notes in Mathematics, {\bf 1735} (Springer, New York, 2000). 

\bibitem{MON}  E. Hern\'andez, A. J\'auregui, A. Mondrag\'on, 
Degeneracy of resonances in a double barrier potential, J. Phys. A: Math. Gen., {\bf 33}, 4507-4523 (2000). 

\bibitem{MON1} E. Hern\'andez, A. J\'auregui, A. Mondrag\'on, 
Energy eigenvalue surfaces close to a degeneracy of unbound states: Crossings and anticrossings of energies and widths, Phys. Rev. E, {\bf 72}, 026221 (2005).


\bibitem{AGML} J.J. Alvarez, M. Gadella, L.P. Lara, F.H.  Maldonado Villamizar, Unstable quantum oscillator with point interactions: Maverick resonances, antibound states and other surprises, Phys. Lett. A, {\bf 377}, 2510Ð2519 (2013). 

\bibitem{VER1}  T. Vertse, R.G. Lovas, A. R\'acz, P. Salamon,  The use of Ixaru's method in locating the poles of the S-matrix in strictly finite-range potentials, AIP Conf. Proc., {\bf 1479}, 1216-1219 (2012). 

\bibitem{VER2} T. Vertse, R.G. Lovas, P. Salamon,  A. R\'acz, Poles of the S-matrix in Woods-Saxon and Salamon-Vertse potentials, AIP Conf. Proc., {\bf 1491}, 113-116 (2012). 

\bibitem{CHIS} D.J. Fern\'andez C., Supersymmetric quantum mechanics, AIP Conference Proceedings, {\bf 1287}, 3-36 (2010). 

\bibitem{BOG1} B. Mielnik, Factorization method and new potentials with the oscillator spectrum, J. Math. Phys., {\bf 25}, 3387-3389 (1984). 

\bibitem{oscar1} N. Fern\'andez-Garc\'ia and O. Rosas-Ortiz, Gamow-Siegert functions and Darboux-deformed short range potentials, Ann. Phys., {\bf 323}, 1397-1414 (2008).

\bibitem{oscar2} N. Fern\'andez-Garc\'ia and O. Rosas-Ortiz, Optical potentials using resonance states in supersymmetric quantum mechanics, J. Phys. Conf. Ser., {\bf 128}, 012042 (2008).

\bibitem{oscar3} N. Fern\'andez-Garc\'ia and O. Rosas-Ortiz, Rectangular potentials in a semi-harmonic background: spectrum, resonances and dwell time, SIGMA, {\bf7}, 044-063 (2011).

\bibitem{AS} M. Abramovitz, I.A. Stegun, {\it Handbook of Mathematical Functions} (Dover, New York, 1965). 

\bibitem{B} L.J. Boya, Quantum mechanical scattering in one dimension,  
Rivista Nuovo Cimento, {\bf 31}, 75-139 (2008). 

\bibitem{Alhassid} Y. Alhassid, F. G\"ursey and F. Iachello, 
{Potential scattering, transfer matrix and group theory},  Phys. Rev. Lett., {\bf 50}, 873-876 (1983). 

\bibitem{Guerrero} J. Guerrero, 
{A group-theoretical derivation of the S-matrix for
the P\"oschl-Teller potentials},  J. Phys. Conf. Ser., {\bf 237}, 012012 (2010). 

\bibitem{Flugge} S. Fl\"ugge,  {\it Practical Quantum Mechanics} (Springer, Berlin, 1999). 

\bibitem{KN} \c{S}. Kuru, J. Negro,  Dynamical algebras for P\"oschl-Teller Hamiltonian hierarchies, Ann. Phys. (NY), {\bf 324}, 2548-2560 (2009). 

\bibitem{MBG} R. de la Madrid, A. Bohm, M. Gadella, Rigged Hilbert Space treatment of the continuous spectrum, Fortschr. Phys., {\bf 50}, 185-216 (2002). 

\bibitem{SJ1}  J.A. Calzada, \c{S}. Kuru, J. Negro, M.A. del Olmo, Dynamical algebras of general two-parametric P\"oschl-Teller Hamiltonians, Ann. Phys., {\bf 327}, 808-822 (2012). 

\bibitem{SJ2}  \c{S}. Kuru, J. Negro, 
Classical spectrum generating algebra of the Kepler-Coulomb system and action-angle variables, Phys. Lett., {\bf 376}, 260-264 (2012). 

\bibitem{PRON}  M. Gadella, J. Negro, L.M. Nieto, G.P. Pronko, M. Santander, Spectrum generating algebras for the free motion in $S^3$, J. Math. Phys., {\bf 52}, 063509 (2011).

\end{thebibliography}
\end{document}